\documentclass[a4paper,11pt]{article}
\usepackage{jheppub} 
\usepackage[T1]{fontenc} 
\RequirePackage[displaymath]{lineno} 
\usepackage{epsfig}
\usepackage{graphicx}
\usepackage{dcolumn}
\usepackage{bm}
\usepackage{ltablex,booktabs}
\usepackage{overpic}
\usepackage{subfigure}
\usepackage{float}
\usepackage{color}
\usepackage{amsmath}
\usepackage{mathcomp}
\usepackage{mathrsfs}
\usepackage{multirow}
\usepackage{rotating}
\usepackage{amssymb}
\usepackage{gensymb}
\usepackage{amsmath}
\usepackage{tabularx}
\usepackage{epstopdf}
\usepackage{relsize}

\title{\boldmath\texorpdfstring{Determination of $U$-spin breaking parameters with an amplitude analysis of the decay $D^{0}\to K_{\text{S},\text{L}}^{0}\pi^{+}\pi^{-}$}{Determination of $U$-spin breaking parameters with an amplitude analysis of the decay D0 to KSL0 pi+ pi-}}
\collaborationImg{\includegraphics[width=0.15\textwidth, angle=90]{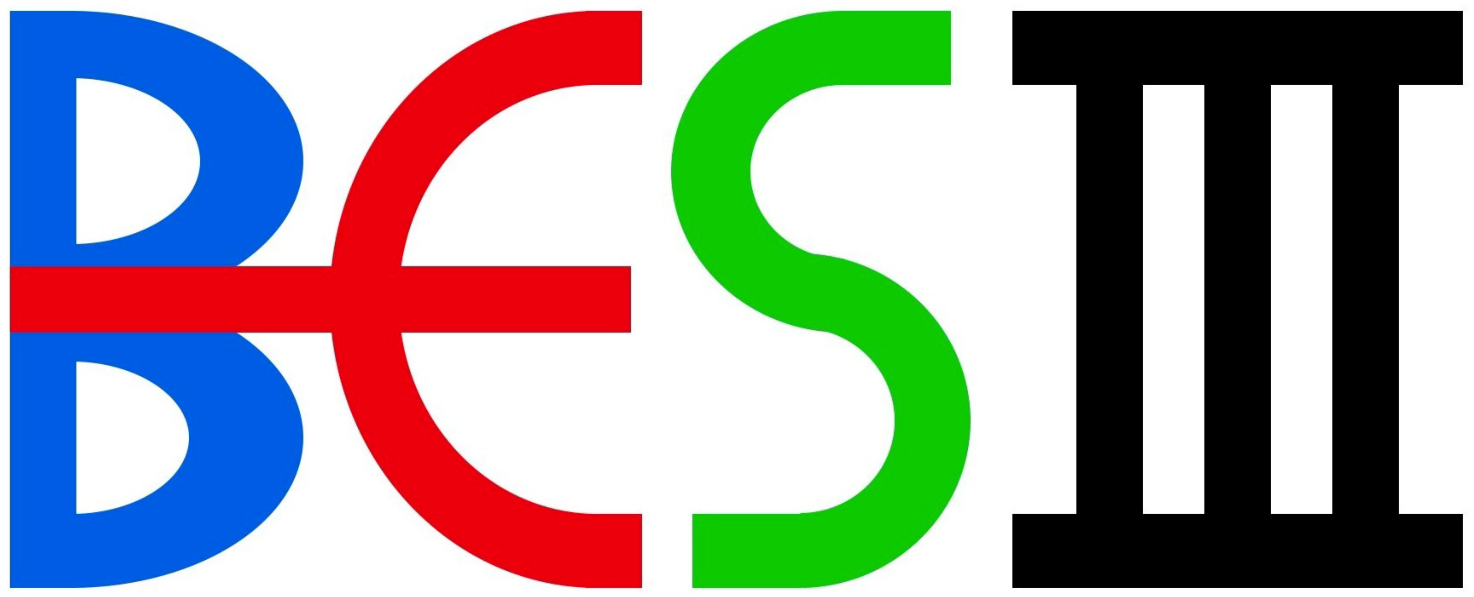}}
\collaboration{BESIII Collaboration}

\abstract{
We perform a simultaneous amplitude analysis of the decays $D^0 \to K_{\text{S}}^{0}\pi^{+}\pi^{-}$ and $D^0 \to K_{\text{L}}^{0}\pi^{+}\pi^{-}$ using an $e^+e^-$ collision data sample, corresponding to an integrated luminosity of $20.3\,\text{fb}^{-1}$, collected at the centre-of-mass energy of $3.773\,\text{GeV}$ with the BESIII detector.
This study, based on a significantly larger dataset than that used in previous work, presents an updated measurement of the complex $U$-spin breaking parameters ($\hat{\rho}$) associated with various $CP$ eigenstate resonant modes mediating the three-body decays.
The magnitudes of the $\hat{\rho}$ parameters range from $0.62$ to $4.45$, indicating substantial $U$-spin symmetry breaking.
We also report the resonant fit fraction and obtain a model prediction for the average strong-phase parameters over regions of phase space for both decay modes.
}

\arxivnumber{}
\begin{document}
\maketitle
\flushbottom

\section{Introduction}
\label{Intro}
The decay $D^0\to K_{\text{S},\text{L}}^0\pi^+\pi^-$ plays a central role in the study of flavour dynamics in three-body charm decays. 
Its rich Dalitz plot (DP) structure exhibits strong interference among intermediate resonances, 
providing sensitivity to strong phases, amplitude dynamics, and flavour-symmetry breaking effects across the phase space. 
This strong-phase structure provides essential input for model-independent determinations of the Cabibbo-Kobayashi-Maskawa (CKM) angle $\gamma$ in 
$B^\pm\to DK^\pm$ decays~\cite{PRD_82_112006,PRL_124_241802,PRD_101_112002,PRD_102_052008,shenwenhan}, as well as for studies of charm mixing through strong-phase-dependent interference effects~\cite{charm_mixing}.
In particular, the interference between Cabibbo-favoured (CF) and doubly Cabibbo-suppressed (DCS) amplitudes, 
together with neutral kaon mixing, induces an asymmetry between the 
$K_\text{S}^0\pi^+\pi^-$ and $K_{\text{L}}^0\pi^+\pi^-$ final states~\cite{Wang:2017ksn,Yu:2017oky}. 

This asymmetry can be characterised in terms of the $U$-spin breaking parameter $\hat{\rho}$~\cite{IIB_HY}, defined through
\begin{equation}
\mathcal{R}(D,\pi\pi) = \frac{\mathcal{B}(D \to K^0_\text{S} \pi^+\pi^-) - \mathcal{B}(D \to K^0_{\text{L}} \pi^+\pi^-)}{\mathcal{B}(D \to K^0_\text{S} \pi^+\pi^-) + \mathcal{B}(D \to K^0_{\text{L}} \pi^+\pi^-)} \simeq 2 \tan^2\theta_C\, \mathrm{Re}\bigl[ \hat{\rho} \bigr],
\label{eq:Rdef}
\end{equation}
where $\theta_C$ is the Cabibbo angle and $\mathcal{B}(X)$ denotes the branching fraction of the decay mode $X$.
In the limit of exact $U$-spin symmetry, one expects $\hat{\rho}=1$, 
and deviations from unity provide sensitivity to flavour-symmetry breaking effects in these decays. 
The real part of $\hat{\rho}$ governs the charge-parity ($CP$) conserving $K_{\text{S}}^0$-$K_{\text{L}}^0$ rate asymmetry, while the imaginary part gives observables sensitive to $CP$ violation through its interplay with strong-phase differences and neutral kaon mixing. 
Since different intermediate resonances receive different CF and DCS contributions across the DP, 
resonance-dependent determinations of $\hat{\rho}$ provide detailed information on 
flavour-symmetry breaking in three-body charm decay dynamics. 
Several theoretical studies have highlighted the sensitivity of the 
$K_\text{S}^0$-$K_{\text{L}}^0$ asymmetry to $U$-spin breaking, strong phases, 
and CF-DCS interference effects~\cite{Wang:2017ksn,Yu:2017oky,Cheng:2024hdo}.

Experimentally, strong-phase variations in these decays have been studied using quantum-correlated 
$D\bar D$ data at the $\psi(3770)$ resonance, where the system forms a coherent antisymmetric state. 
These measurements are encoded in the binned coefficients $(c_i,s_i)$ and $(c_i^{\prime},s_i^{\prime})$, 
which provide essential input for charm mixing and model-independent determinations of the CKM angle $\gamma$~\cite{PRD_82_112006,PRL_124_241802,PRD_101_112002,PRD_102_052008,shenwenhan}. 
However, these measurements do not determine the underlying resonance-dependent $U$-spin breaking responsible for the differences between the $K_{\text{S}}^0\pi^+\pi^-$ and $K_{\text{L}}^0\pi^+\pi^-$ decay amplitudes. A first attempt to extract the corresponding $U$-spin breaking parameter $\hat{\rho}$ was performed in a preliminary BESIII analysis based on $2.93\,\mathrm{fb}^{-1}$ of data~\cite{BESIII:2022qvy}, but was limited by the available statistics and a simplified amplitude model.
With the full dataset corresponding to an integrated luminosity of $20.3\,\mathrm{fb}^{-1}$~\cite{CPC_37_123001,CPC_48_123001}, a more complete treatment of the decay dynamics becomes possible, 
allowing a simultaneous amplitude analysis of the $D^0\to K_{\text{S}}^0\pi^+\pi^-$ and $D^0\to K_{\text{L}}^0\pi^+\pi^-$ decays.

In this work, we construct a unified amplitude framework and extract the resonance-dependent complex $U$-spin breaking parameters $\hat{\rho}$, together with the $S$-wave contribution, directly from data. The analysis provides a consistent amplitude-level description of $U$-spin breaking effects and yields model predictions for the strong-phase coefficients $(c_i,s_i)$ and $(c_i^{\prime},s_i^{\prime})$. 
While these predictions do not supersede the model-independent measurements from quantum-correlated $D\bar D$ data, 
the resulting amplitude model may be useful for evaluating model-dependent systematic effects in such measurements, for example through studies of efficiency corrections across the Dalitz plot. 
The primary outcome of this work is an updated measurement of the resonance-dependent $U$-spin breaking parameters, 
which provides quantitative input for understanding flavour-symmetry breaking in three-body charm decays. 

This paper is organised as follows. 
Section~\ref{sec:detector_dataset} describes the BESIII detector and the data sample. 
Section~\ref{ST-selection} presents the event selection criteria. 
Section~\ref{sec:Amplitude-Analysis} introduces the simultaneous amplitude-analysis framework for 
$D^0\to K_\text{S}^0\pi^+\pi^-$ and $D^0\to K_{\text{L}}^0\pi^+\pi^-$ decays, including the parametrisation of the decay amplitudes and the treatment of the resonance-dependent $U$-spin breaking parameters $\hat{\rho}$. 
Section~\ref{sec:systematics} discusses the evaluation of systematic uncertainties. 
Section~\ref{sec:summary} summarises the results and their implications.
\section{Detector and data sample}
\label{sec:detector_dataset}

The BESIII detector~\cite{Ablikim:2009aa} records symmetric $e^+e^-$ collisions delivered by the BEPCII storage ring~\cite{Yu:IPAC2016-TUYA01} in the centre-of-mass (CM) energy range from 1.84 to 4.95~GeV, achieving a peak luminosity of $1.1 \times 10^{33}\,\text{cm}^{-2}\,\text{s}^{-1}$ at $\sqrt{s} = 3.773\,\text{GeV}$. BESIII has collected large data samples~\cite{Ablikim:2019hff, EcmsMea, EventFilter}. The cylindrical core of the BESIII detector covers 93\% of the full solid angle and comprises a helium-based multilayer drift chamber (MDC), a plastic scintillator time-of-flight (TOF) system, and a CsI(Tl) electromagnetic calorimeter (EMC), all enclosed in a superconducting solenoidal magnet providing a 1.0~T magnetic field. The solenoid is supported by an octagonal flux-return yoke with resistive plate counter muon identification modules interleaved with steel layers. The charged-particle momentum resolution at $1\,\text{GeV}/c$ is $0.5\%$, and the $\text{d}E/\text{d}x$ resolution is $6\%$ for electrons from Bhabha scattering. The EMC measures photon energies with a resolution of 2.5\% (5\%) at 1 GeV in the barrel (end cap) region. The time resolution of the barrel TOF is 68~ps, while that in the end cap TOF is 110 ps. The end cap TOF system was upgraded in 2015 using multigap resistive plate chamber technology, providing a time resolution of 60~ps, which benefits 86\% of the data used in this analysis~\cite{etof1,etof2,etof3}.

Monte Carlo (MC) simulated samples, produced with a {\sc{geant4}}-based software package~\cite{geant4}, are used to determine detection efficiencies and estimate backgrounds. The simulation includes a detailed geometric description of the BESIII detector, its response, and models the beam-energy spread and initial-state radiation in $e^+e^-$ annihilations using the {\sc kkmc} generator~\cite{KKMC1,KKMC2}. The inclusive MC sample incorporates the production of $D\bar{D}$ pairs (with quantum-coherence effects for neutral $D^0 \bar D^0$ channels), non-$D\bar{D}$ decays of the $\psi(3770)$, initial-state radiation production of $J/\psi$ and $\psi(3686)$ states, and continuum processes as implemented in {\sc kkmc}~\cite{KKMC1,KKMC2}.

This analysis uses a data sample corresponding to an integrated luminosity of $20.3\,\mathrm{fb}^{-1}$, collected at the CM energy $\sqrt{s} = 3.773\,\mathrm{GeV}$. The $\psi(3770)$ predominantly decays to $D^+D^-$ or $\bar{D}^0 D^0$ pairs, with heavily suppressed non-$D\bar{D}$  decays, making it an ideal environment for studying $D$ meson decays via the double-tag (DT) technique~\cite{DTmethod}. In this method, a single-tag (ST) candidate is defined by reconstructing a single $\bar D^0$ meson through one of three hadronic decay modes: $\bar D^0 \to K^+\pi^-$, $\bar D^0 \to K^+\pi^-\pi^0$, and $\bar D^0 \to K^+\pi^-\pi^-\pi^+$. A DT candidate requires the reconstruction of both $D^0$ and $\bar D^0$ mesons, where the $D^0$ decays to the signal decay $D^0 \to K^0_{\text{S},\text{L}} \pi^+ \pi^-$ and the $\bar D^0$ decays to one of the ST modes.

All particle decays are modelled with {\sc evtgen}~\cite{EVTGEN1,EVTGEN2}, using branching fractions (BFs) taken from the Particle Data Group (PDG)~\cite{PDG} when available, and remaining unknown decays are generated with {\sc lundcharm}~\cite{LUNDCHARM1,LUNDCHARM2}. Final-state radiation from charged final-state particles is included using {\sc photos}~\cite{PHOTOS}.
A phase-space (PHSP) MC sample is generated with a uniform distribution for the decays $D^0 \to K_\text{S}^0\pi^+\pi^-$ and $D^0 \to K_{\text{L}}^0\pi^+\pi^-$. This sample is used to determine the detection efficiency function described in Sec.~\ref{sec:fitmethod} and to calculate the normalisation integral required for determining amplitude-model parameters in the data fit. Charge-conjugate processes are implied throughout unless stated otherwise in this work.
\section{Event selection}
\label{ST-selection}

The $D^0$ candidates are constructed from individual $K^{\pm}$, $\pi^{\pm}$, $\pi^0$, and $K^0_\text{S}$ candidates with the following selection criteria.  Additional requirements used in the amplitude analysis are discussed in Sec.~\ref{sec:pwa-select}. 
 
All charged tracks detected in the MDC must satisfy $\vert\!\cos\theta\vert<0.93$, where $\theta$ is the polar angle with respect to the $z$-axis, the symmetry axis of the MDC. For charged tracks not originating from $K_\text{S}^0$ decays, the distance of closest approach to the interaction point is required to be less than 1\,cm in the transverse plane, and less than 10\,cm along the $z$-axis. 

Particle identification (PID) for charged tracks combines the ${\mathrm d}E/{\mathrm d}x$ measurement from the MDC with the time-of-flight measurement from the TOF detector to define the likelihood function $\mathcal{L}(h)~(h=K,\pi)$ for each hadron hypothesis $h$. Charged kaons and pions are identified by comparing the kaon and pion likelihoods, $\mathcal{L}(K)>\mathcal{L}(\pi)$ for kaons and $\mathcal{L}(\pi)>\mathcal{L}(K)$ for pions. 

Photon candidates are identified using isolated showers in the EMC. The deposited energy of each shower must be more than 25~MeV in the barrel region ($|\cos\!\theta|< 0.80$) and more than 50~MeV in the end-cap region ($0.86 <|\cos\!\theta|< 0.92$). To exclude showers originating from charged tracks, the angle between the EMC shower and the position of the closest charged track at the EMC must be greater than $10^{\circ}$ as measured from the interaction point. To suppress electronic noise and showers unrelated to the event, the difference between the EMC time and the event start time is required to be within $[0, 700]\,\mathrm{ns}$.

The $\pi^0$ candidates are formed from photon pairs with invariant masses in the range $[0.115, 0.150]~\mathrm{GeV}/c^2$, which corresponds to about three times the invariant mass resolution. Moreover, to achieve adequate mass resolution, at least one of the two photons is required to be detected in the barrel EMC. A kinematic fit that constrains the $\gamma\gamma$ invariant mass to the known $\pi^{0}$ meson mass~\cite{PDG} is performed to improve the mass resolution. The $\chi^2$ of the kinematic fit is required to be less than 50. 

Each $K_{\text{S}}^0$ candidate is reconstructed from two oppositely charged tracks satisfying \mbox{$|V_{z}| < 20\,\text{cm}$}. The two charged tracks are assigned as $\pi^+\pi^-$ without imposing PID. They are constrained to originate from a common vertex and are required to have an invariant mass within $|M_{\pi^{+}\pi^{-}} - m_{K_{\text{S}}^{0}}| < 12\,\mathrm{MeV}/c^{2}$, where $m_{K_{\text{S}}^{0}}$ is the known $K^0$ meson mass~\cite{PDG}. The decay length of the $K^0_\text{S}$ candidate is required to be greater than twice its resolution.

In this work, three tag modes $\bar{D}^{0} \to K^{+}\pi^{-}$, $\bar{D}^{0} \to K^{+}\pi^{-}\pi^{0}$ and $\bar{D}^{0} \to K^{+}\pi^{-}\pi^{-}\pi^{+}$ are used. The tagged $\bar{D}^{0}$ mesons are selected using two variables, the energy difference and the beam-constrained (BC) mass
\begin{equation}
  \begin{aligned}
    \Delta{E} &= E_D-E_{\text{beam}},
    M_{\text{BC}} &= \sqrt{E_{\text{beam}}^2-|\vec{p}_{D}|^2},\\  
  \end{aligned}	
\end{equation}
where $E_{\text{beam}}$ is the beam energy, $\vec{p}_D$ and $E_D$ are the momentum and energy of the $\bar{D}^{0}$ candidate in the $e^{+}e^{-}$ rest frame. Signal $D^0$ candidates are expected to peak at the known $D^0$ meson mass~\cite{PDG} in $M_{\text{BC}}$ and with zero in $\Delta{E}$. The boundaries of $\Delta E_{\text{tag}}$ requirements are listed in Table~\ref{tagdeltaE}. If there are multiple signal candidates in an event for the decay $D^0 \to K_\text{S}^0\pi^+\pi^-$, the best candidate with the minimum sum of $\Delta{E}$ squared of two $D$ mesons ($\Delta{E}_{\text{tag}}^2+\Delta{E}_{\text{sig}}^2$) is chosen. Then we require $1.860<M_{\text{BC}}^{\text{sig}}<1.871\,\mathrm{GeV}/c^2$ and $-0.030<\Delta{E_{\text{sig}}}<0.030$~GeV as the signal region.

\begin{table}[hbtp]
    \centering
    \caption{\small Requirements of $\Delta E_{\text{tag}}$ and $M_{\text{BC}}^{\text{tag}}$ for different tag modes.}
    \begin{tabular}{lcc}
    \hline
    \hline
    Tag mode & $\Delta E_{\text{tag}}$ (GeV) & $M_{\text{BC}}^{\text{tag}}$ ($\mathrm{GeV}/c^2$) \\
    \hline
    $\bar{D}^{0} \to K^{+}\pi^{-}$ & $(-0.027, 0.027)$ &$ (1.859, 1.873) $\\
    $\bar{D}^{0} \to K^{+}\pi^{-}\pi^{0}$ &$ (-0.062, 0.049)$ & $(1.859, 1.873) $\\
    $\bar{D}^{0} \to K^{+}\pi^{-}\pi^{-}\pi^{+}$ & $(-0.026, 0.024) $& $(1.859, 1.873)$ \\
    \hline
    \hline
    \end{tabular}
    \label{tagdeltaE}
\end{table}

In the selection of $D^{0} \to K_{\text{L}}^0\pi^+\pi^-$ decays, after a tag side $\bar{D}^0$  candidate is identified, we reconstruct the decay $D^0 \to K_{\text{L}}^0 \pi^+\pi^-$ recoiling against the tag side, requiring only two opposite charged tracks identified as a $\pi^+\pi^-$ pair with the same selection criteria as the tag side.
The residual four-momentum in the detector, after reconstructing all the charged tracks on the tag side and both the pion tracks on the signal side, is called missing momentum and is identified as a $K_{\text{L}}^0$ candidate. If there are multiple signal candidates in an event, the candidate with the smallest $\Delta{E}_{\text{tag}}$ is retained. To avoid a misidentification of showers produced in the interactions of $K_{\text{L}}^0$ in the EMC as photons originating from a $\pi^0$ decay, photons assigned to the $\pi^0$ subsystem are additionally required to have an opening angle larger than $15^\circ$ to the $K_{\text{L}}^0$ trajectory. To suppress dominant peaking background from $D^0 \to K_\text{S}^0\pi^+\pi^-$ decays with $K_\text{S}^0\to \pi^0\pi^0$, the total energy of extra showers that are situated away from the $K_{\text{L}}^0$ direction with an opening angle larger than $15^\circ$, $E_{\gamma}^{\text{extra}}$, is required to be less than 0.235~GeV. To further suppress the background, the missing-mass-squared variable is defined as
\begin{equation}
    {{M}_{\text{miss}}^{2}} = (\sqrt{s}/2-E_{\pi^-}-E_{\pi^+})^{2} - \left|\vec{p}_{\text{tag}}+\vec{p}_{\pi^-}+\vec{p}_{\pi^+}\right|^{2},
\end{equation}
and is required to satisfy $0.202<M_{\text{miss}}^{2}<0.296\,\text{GeV}^2/c^4$. In the above equation, ($E_{\pi^{\pm}},\vec{p}_{\pi^{\pm}}$) denote the four-momentum of $\pi^{\pm}$ candidates on the signal side and $\vec{p}_{\text{tag}}$ is the total momentum of the ST $\bar{D}^0$.

\section{Amplitude analysis}
\label{sec:Amplitude-Analysis}
\subsection{Background study and signal yields extraction}
\label{sec:pwa-select}

The main peaking backgrounds for the $D^0 \to K_\text{S}^0 \pi^+\pi^-$ decay are from the $D^0 \to K_\text{S}^0 K_\text{S}^0$ and $D^0 \to \pi^+\pi^-\pi^+\pi^-$ processes. 
The main peaking background for the $D^0 \to K_{\text{L}}^0 \pi^+\pi^-$ decay comes from $D^0 \to K_\text{S}^0\pi^+\pi^-$ decays when the $K_\text{S}^0$ is missing.
To suppress the $K_\text{S}^0 \to \pi^+\pi^-$ background, a $K_\text{S}^0$ veto is applied for the reconstructed $\pi^+\pi^-$ pairs. We perform a vertex fit on the $\pi^+\pi^-$ pairs and require the decay length significance $L/\sigma$ to be less than 2. 

To ensure that all candidates fall within the PHSP boundary, a six-constraint (6C) kinematic fit is performed, which includes constraints of four-momentum conservation, the $D^0$ (signal) and $K_\text{S}^0$ meson masses for the $D^0 \to K_\text{S}^0\pi^+\pi^-$ decay. For the $D^{0}\to K_{\text{L}}^{0}\pi^{+}\pi^{-}$ decay, the missing $K^{0}_{\text{L}}$ information results in a kinematic configuration with 6C and 4 free parameters (the four-momentum of the $K_{\text{L}}^{0}$). The four unknown components of the $K_L^0$ four-momentum reduce the effective number of constraints from six to two: the $D^0$ (signal) and $K_L^0$ meson mass constraints. The updated four-momenta will be used in the subsequent amplitude analysis.

To estimate the signal purity and signal yields, we perform a maximum likelihood fit to the distributions of $M_{\text{BC}}^{\text{tag}}$ versus $M_{\text{BC}}^{\text{sig}}$ for the $D^0 \to K_\text{S}^0 \pi^+\pi^-$ mode and ${M}_{\text{miss}}^{2}$ for the $D^0 \to K_{\text{L}}^0 \pi^+\pi^-$ mode.

The signal events with both the tag and the signal
side correctly reconstructed concentrate around
$M_{\text{BC}}^{\text{sig}} = M_{\text{BC}}^{\text{tag}} = m_{D^0}$, where $m_{D^0}$ is
the known ${D^0}$ meson mass. 

\begin{figure}[htbp] 
  \centering
  \includegraphics[width=0.8\linewidth]{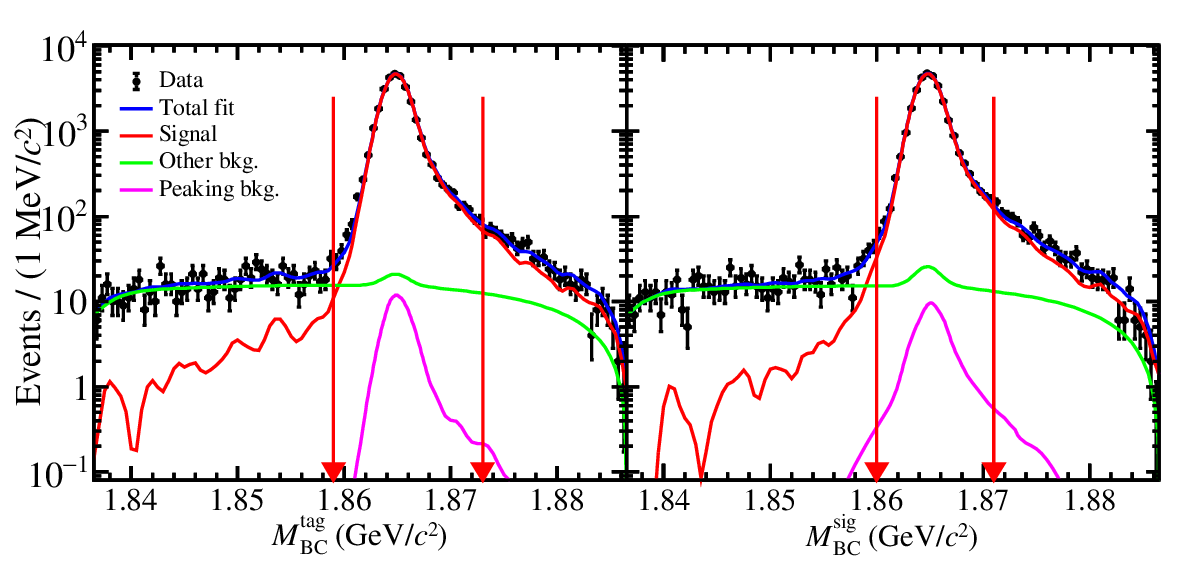}\\
  \includegraphics[width=0.8\linewidth]{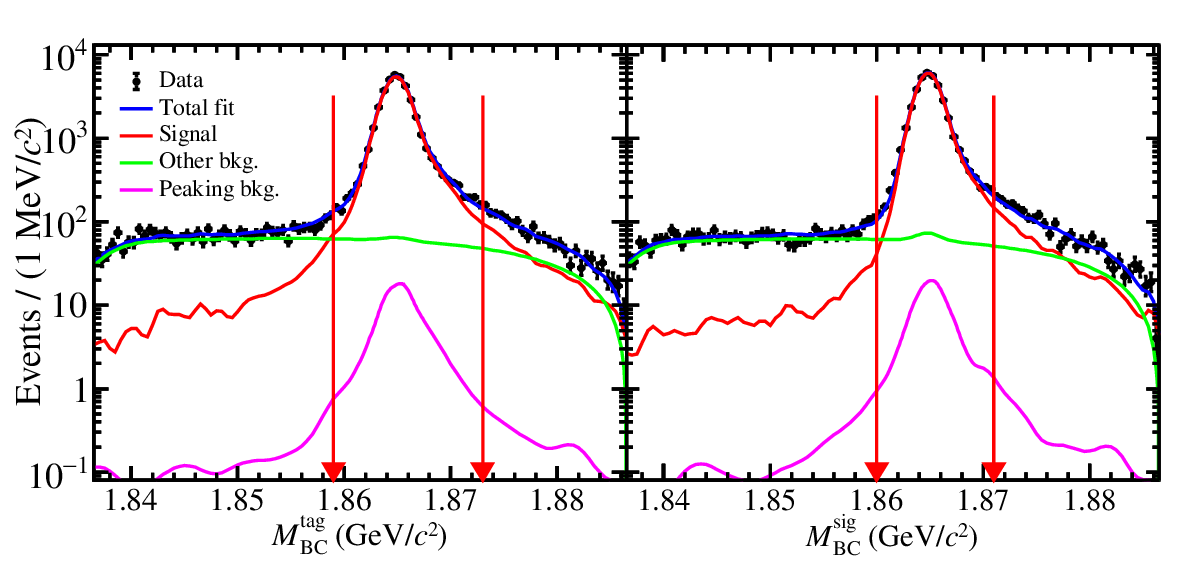}\\
  \includegraphics[width=0.8\linewidth]{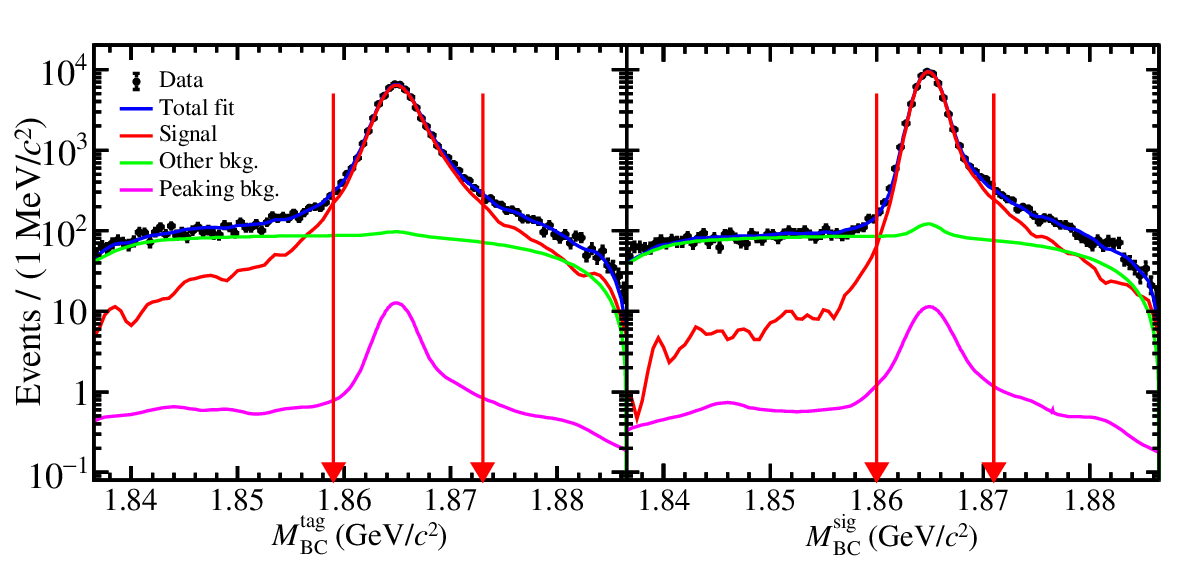}\\
  \caption{ Projections of the 2D fit on (left) $M_{\text{BC}}^{\text{tag}}$ and (right) $M_{\text{BC}}^{\text{sig}}$. Data points with error bars are shown. The blue curves show the total fit, the red curves represent the signal component, the pink curves show the peaking background (bkg.), and the green curves represent the other bkg. (BG-I, BG-II and BG-III). The red arrows indicate the $M_{\text{BC}}$ signal region. From top to bottom, the rows correspond to candidate events for $D^0 \to K_\text{S}^0\pi^+\pi^-$ versus $\bar{D}^0 \to K^{+}\pi^-$, $\bar{D}^0 \to K^{+}\pi^-\pi^0$ and $\bar{D}^0 \rightarrow K^{+} \pi^{-} \pi^{-} \pi^{+}$. }
  \label{fig:KSpipi_2Dfit}
\end{figure}

Besides signal events, we identify four kinds of backgrounds. Candidates with correctly
reconstructed $D^0$ (or $\bar{D}^0$) and incorrectly reconstructed $\bar{D}^0$ (or $D^0$) are
denoted as BG-I, which appear around the lines $M_{\text{BC}}^{\text{sig}} = m_{D^0}$ or
$M_{\text{BC}}^{\text{tag}} = m_{D^0}$. Other candidates appearing along the
diagonal (mispartitioning continuum) arise from
decay modes sharing the same final state particle combination with the signal~(BG-II).
The remaining background events mainly come from candidates reconstructed incorrectly on both sides (BG-III). In addition, the main peaking backgrounds for the $D^0 \to K_\text{S}^0 \pi^+\pi^-$ signal decay are from the $D^0 \to K_\text{S}^0 K_\text{S}^0$ and $D^0 \to \pi^+\pi^-\pi^+\pi^-$ processes (BG-IV).
Here we list the probability density functions (PDFs) for the corresponding components~\cite{2dfitPDFs,2Dfit-cleo} in the fit:
\begin{itemize}
    \item Signal: $s(x, y)$,
    \item BG-I: $b(x) \cdot c_y\left(y; E_{\text{beam}}, \xi_y\right) + b(y) \cdot c_x\left(x; E_{\text{beam}}, \xi_x\right)$,
    \item BG-II: $c_z\left(z; \sqrt{2} E_{\text{beam}}, \xi_z\right) \cdot g\left(k; 0, \sigma_k\right)$,
    \item BG-III: $c_x\left(x; E_{\text{beam}}, \xi_x\right) \cdot c_y\left(y; E_{\text{beam}}, \xi_y\right)$,
    \item BG-IV: $p(x,y)$,
\end{itemize}
where $x=M_{\text{BC}}^{\text{tag}}$, $y=M_{\text{BC}}^{\text{sig}}$, $z=(x+y) / \sqrt{2}$, and $k = (x-y) / \sqrt{2}$. The signal shape $s(x, y)$ is described by the two-dimensional (2D) MC-simulated shape of $D^{0} \bar{D}^{0}$ convolved with two independent Gaussian functions. For BG-I, the function $b(x)$ is described by the one-dimensional (1D) MC-simulated shape of $D^{0}\left(\bar{D}^{0}\right)$ convolved with a Gaussian. The parameters of the Gaussian functions are obtained by a 1D fit to the $M_{\text{BC}}$ distributions on the signal and tag sides and are fixed in the 2D fit. The BG-II fit component is given by an ARGUS function in the diagonal axis multiplied by a Gaussian in the anti-diagonal axis. The function $c_f\left(f ; E_{\text{beam}}, \xi_f\right)$ is the ARGUS function~\cite{argus} defined as

$$
c_f\left(f ; E_{\text{beam}}, \xi_f\right)=A_f f\left(1-\frac{f^2}{E_{\text{beam}}^2}\right)^{\frac{1}{2}} e^{\xi_f\left(1-\frac{f^2}{E_{\text{beam}}^2}\right)},
$$
where $f$ denotes $x, y$, or $z$, $E_{\text{beam}}$ is fixed at 1.8865~GeV, $A_f$ is a normalisation factor, and $\xi_f$ is a fit parameter. The function $g\left(k; 0,\sigma_k\right)$ is a Gaussian distribution with a mean value of zero and a standard deviation $\sigma_k = \sigma_0 \cdot (\sqrt{2} E_{\text{beam}} - z)^p$, where $\sigma_0$ and $p$ are parameters determined by the fit. An ARGUS function gives the BG-III fit component in $M_{\text{BC}}^{\text{sig}}$ multiplied by an ARGUS function in $M_{\text{BC}}^{\text{tag}}$. The inclusive MC sample shape, convolved with a Gaussian function, describes the BG-IV fit component shape. The fit results are shown in Fig.~\ref{fig:KSpipi_2Dfit} and Table~\ref{KSpipi_purity}.

\begin{figure}[htbp] 
  \centering
  \includegraphics[width=0.3\linewidth]{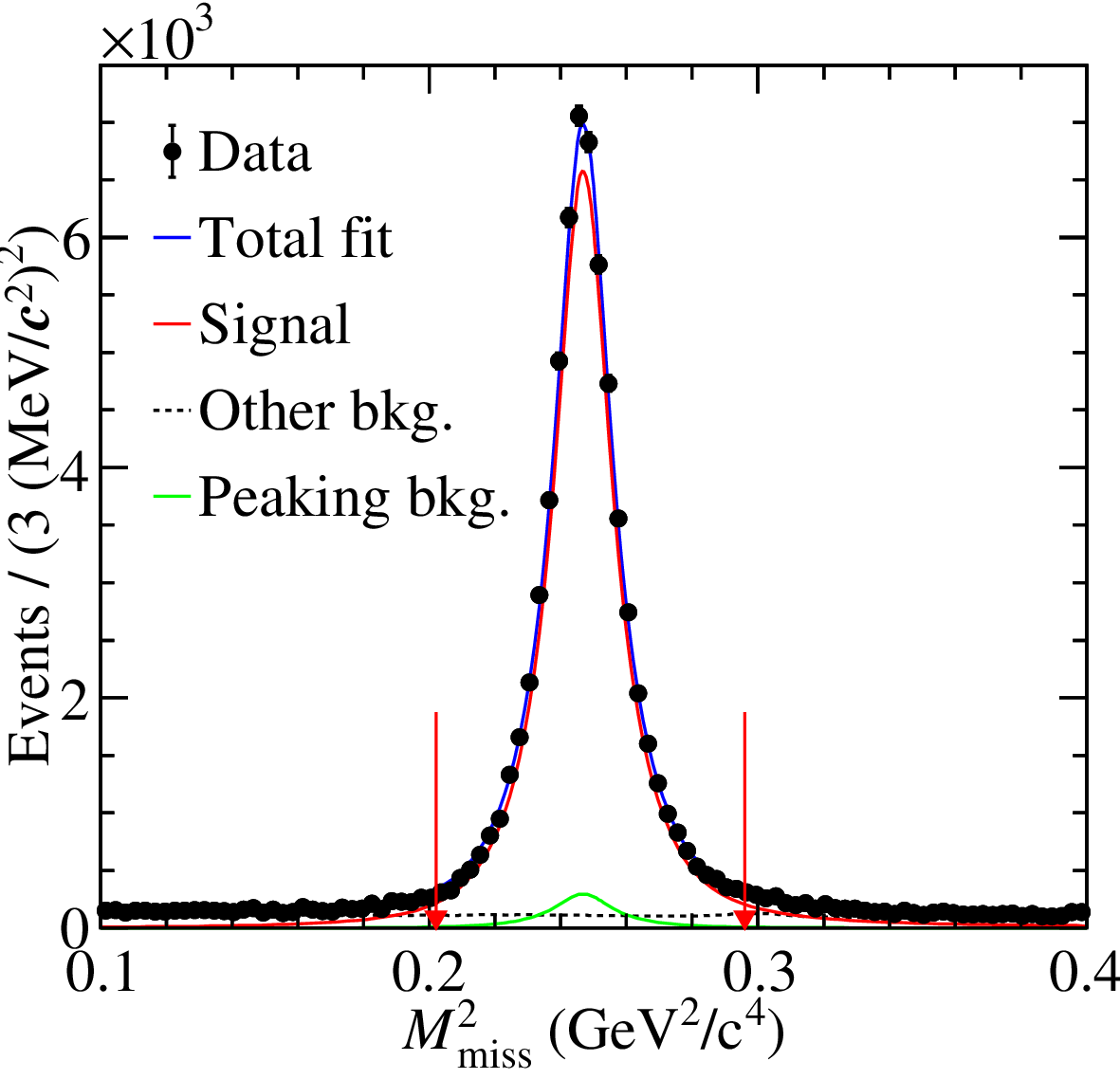}
  \includegraphics[width=0.3\linewidth]{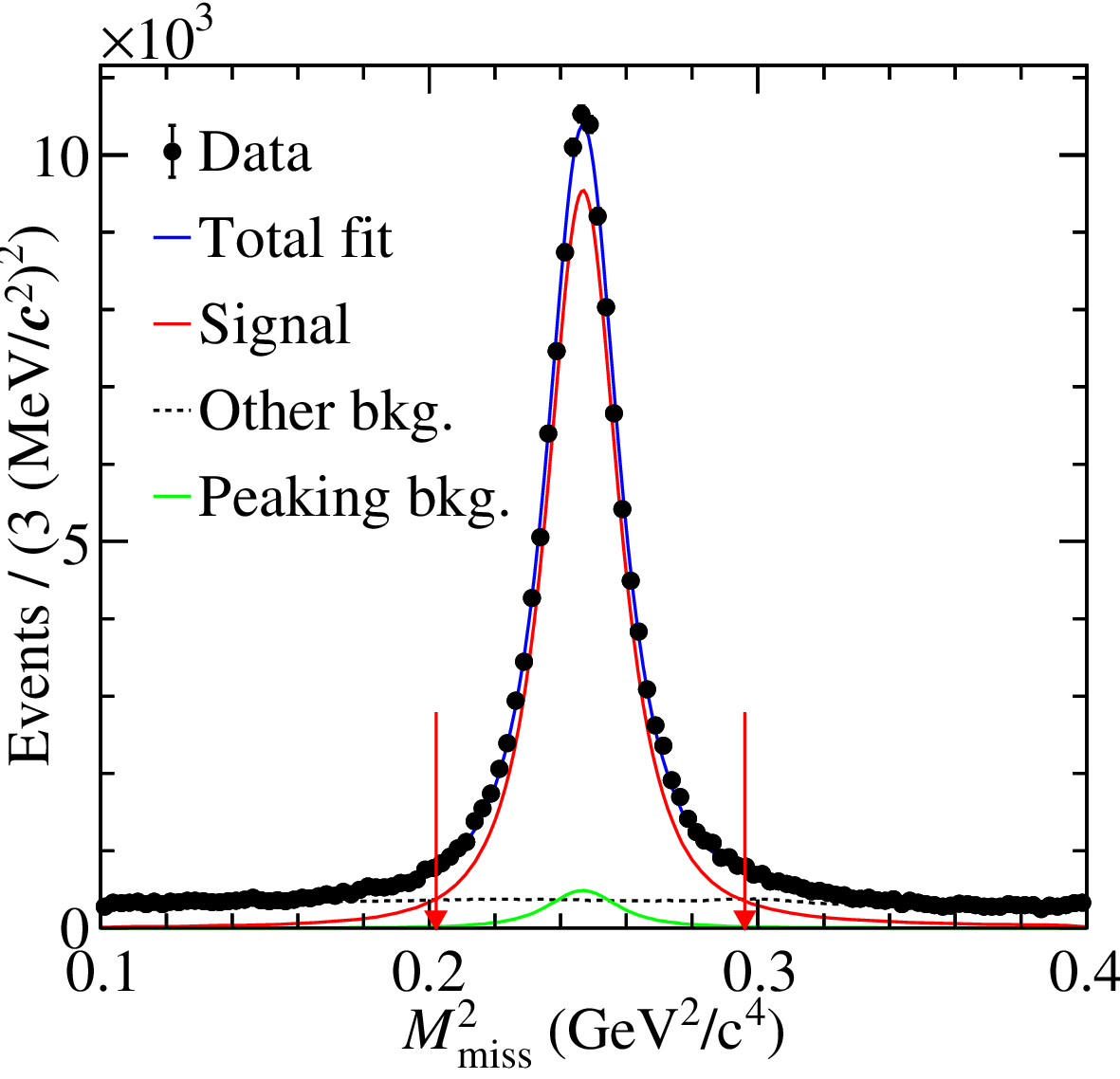}
  \includegraphics[width=0.3\linewidth]{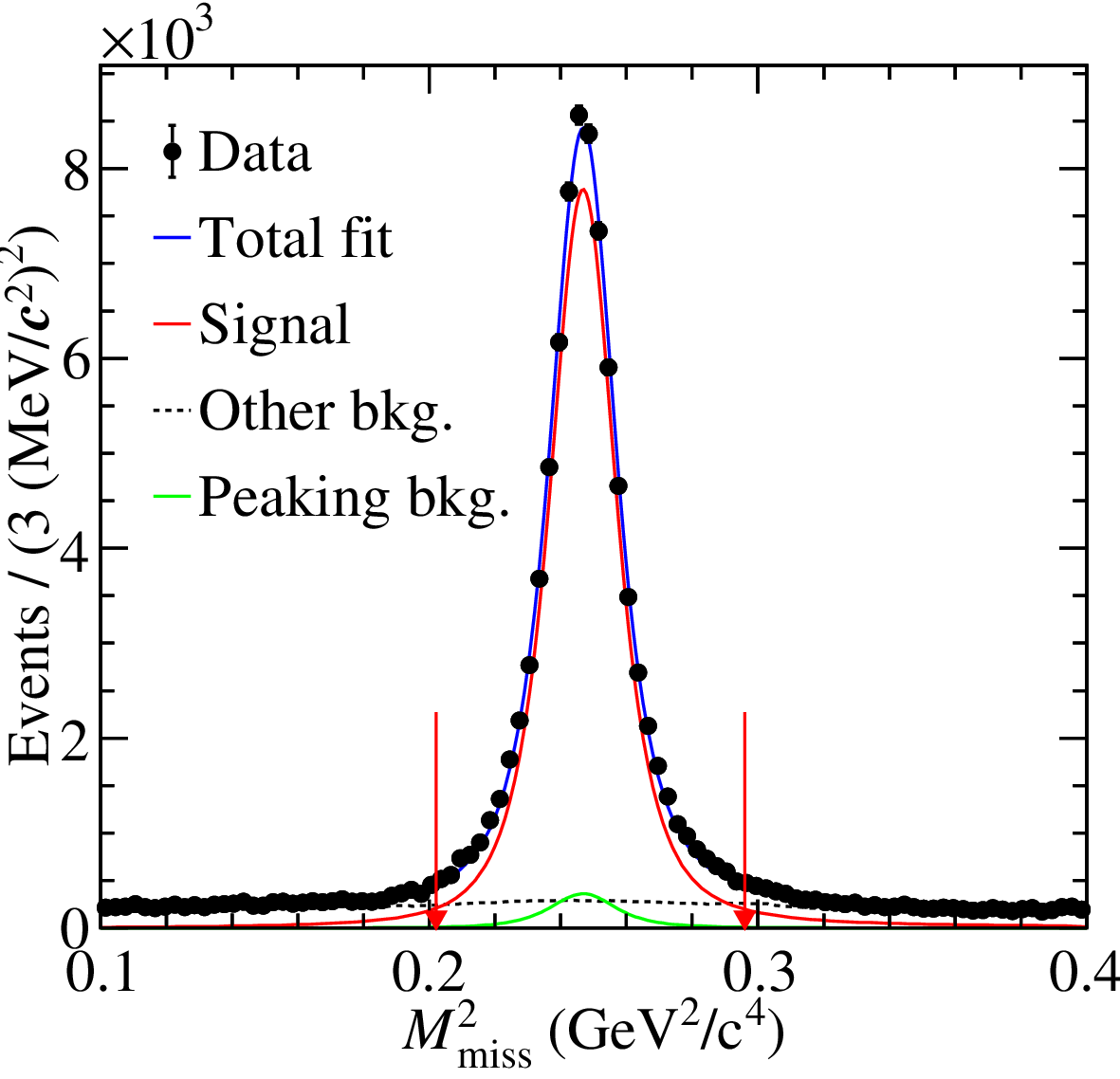}

  \caption{ Projections of the one-dimensional fit on $M_{\text{miss}}^{2}$. The blue curves show the total fit, the red curves represent the signal component, the green curves show the peaking bkg., and the black dashed curves represent the other bkg. The red arrows indicate the $M_{\text{miss}}^{2}$ signal region. From left to right, the panels correspond to candidate events for $D^0 \to K_{\text{L}}^0\pi^+\pi^-$ versus $\bar{D}^0 \to K^{+}\pi^-$, $\bar{D}^0 \to K^{+}\pi^-\pi^0$ and $\bar{D}^0 \rightarrow K^{+} \pi^{-} \pi^{-} \pi^{+}$.}
  \label{fig:KLpipi_1Dfit}
\end{figure}

For the $D^0 \to K_{\text{L}}^0\pi^+\pi^-$ decay, to estimate the signal purity and signal yields, we perform 1D maximum likelihood fits on the $M_{\text{miss}}^{2}$ distributions for the three tag modes separately. The $M_{\text{miss}}^{2}$ distribution is modelled with a signal shape convolved with a Gaussian function, the combinatorial background is modelled by the inclusive MC background shape, and the peaking background shape is identical to the signal shape, while the yield of the peaking background is constrained with a Gaussian prior according to its expected yield derived from the measured BF~\cite{PDG}. The fit results are shown in Fig.~\ref{fig:KLpipi_1Dfit} and Table~\ref{KLpipi_purity}.

\begin{table}[hbtp]
    \centering
    \setlength{\extrarowheight}{2pt}
    \caption{\small Signal yields of $D^0 \to K_\text{S}^0\pi^+\pi^-$ and the corresponding signal purity of the three ST modes in the signal region.}
    \begin{tabular}{l|cc}
    \hline
    \hline
    Tag mode & Signal yield & Signal purity $\left(\omega_{\text{sig}}\right)$ \\
    \hline
    $\bar{D}^0 \rightarrow K^{+} \pi^{-}$ & $29606 \pm 174$ & $(98.5 \pm 0.8) \%$ \\
    $\bar{D}^0 \rightarrow K^{+} \pi^{-} \pi^0$ & $36784 \pm 198$ & $(96.2 \pm 0.7) \%$ \\
    $\bar{D}^0 \rightarrow K^{+} \pi^{-} \pi^{-} \pi^{+}$ & $56966 \pm 245$ & $(96.8 \pm 0.6) \%$ \\
    \hline
    \hline
    \end{tabular}
    \label{KSpipi_purity}
\end{table}

\begin{table}[hbtp]
    \centering
    \setlength{\extrarowheight}{2pt}
    \caption{\small Signal yields of $D^0 \to K_{\text{L}}^0\pi^+\pi^-$ and the corresponding signal purity of the three ST modes in the signal region.}
    \begin{tabular}{l|cc}
    \hline
    \hline
    Tag mode & Signal yield & Signal purity $\left(\omega_{\text{sig}}\right)$ \\
    \hline
    $\bar{D}^0 \rightarrow K^{+} \pi^{-}$ & $58499 \pm 272$ & $(87.6 \pm 0.3) \%$ \\
    $\bar{D}^0 \rightarrow K^{+} \pi^{-} \pi^0$ & $112394 \pm 385$ & $(85.2 \pm 0.3) \%$ \\
    $\bar{D}^0 \rightarrow K^{+} \pi^{-} \pi^{-} \pi^{+}$ & $73110 \pm 311$ & $(84.6 \pm 0.3) \%$ \\
    \hline
    \hline
    \end{tabular}
    \label{KLpipi_purity}
\end{table}

\subsection{Fit method}
\label{sec:fitmethod}

The relative magnitudes and phases of the amplitudes, and the mass and width
of intermediate resonances, are determined by an unbinned maximum-likelihood
fit to the selected data sample. The decay amplitudes are constructed using the covariant-tensor formalism.
The amplitude of the $n^{\text{th}}$ intermediate state ($\mathcal{A}_n$) at position $\textbf{x}$ in the DP is given by
\begin{equation}
  \mathcal{A}_n(\textbf{x}) = P_nS_nF_n^rF_n^D,
\end{equation}
where $S_n$ is the spin factor, $F_n^r$ and $F_n^D$ are the Blatt--Weisskopf barrier factors 
of the intermediate resonance and the $D$ meson, respectively, and $P_n$ is the propagator 
of the intermediate resonance.

Non-trivial effects on the rate of hadronic decays involving pions and neutral kaons, arising from the interference between DCS and CF transitions, were predicted in Ref.~\cite{IIB_HY}. A manifestation of this interference effect is a small difference in the decay rates of processes involving a $K_{\text{S}}^0$ meson versus a $K_{\text{L}}^0$ meson in the final state. For the decays $D^0 \to K_{\text{S}}^0\pi^+\pi^-$ and $D^0 \to K_{\text{L}}^0\pi^+\pi^-$, the amplitude construction can differ due to the interference between CF and DCS sub-transitions, as illustrated in Fig.~\ref{fig:CF_DCS}. In addition to the pure CF and DCS sub-transitions, a mixture of these, where the two pions are produced in a $CP$ eigenstate resonance, is also a viable transition process to the $K_{\text{S}}^0\pi^+\pi^-$ final state. Using the phase convention $CP\left|K^0\right\rangle=-\left|\bar{K}^0\right\rangle$, the partial amplitude for intermediate processes that involve only neutral $CP$-eigenstate resonances $k_{\text{CP}}$, i.e., $D^0 \rightarrow K_{\text{S}}^0(\pi^+\pi^-)_{k_{\text{CP}}}$, can be written as a superposition of $D^0 \rightarrow \bar{K}^0\pi^+\pi^-$ and $D^0 \rightarrow K^0\pi^+\pi^-$ as follows:

\begin{figure*}[htbp] 
  \centering
  \includegraphics[width=12cm]{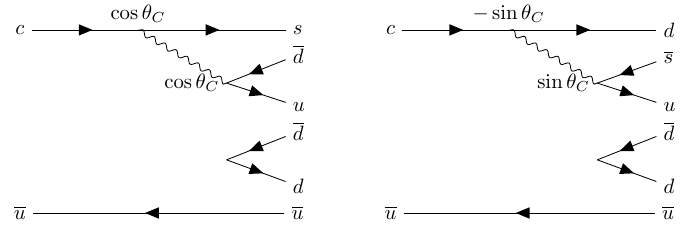}

  \caption{(left) CF $D^0\to \bar{K}^0\pi^+\pi^-$ and (right) DCS $D^0\to K^0\pi^+\pi^-$ decay diagrams.}
  \label{fig:CF_DCS}
\end{figure*}

\begin{equation}
\begin{aligned}
\mathcal{A}\left(D^0 \rightarrow K_{\text{S}}^0(\pi\pi)_{k_{\mathrm{CP}}}\right) & =\frac{1}{\sqrt{2}}\left(\mathcal{A}\left(D^0 \rightarrow \bar{K}^0(\pi^+\pi^-)\right)-\mathcal{A}\left(D^0 \rightarrow K^0(\pi^+\pi^-)\right)\right) \\
& =\frac{1}{\sqrt{2}} \mathcal{A}\left(D^0 \rightarrow \bar{K}^0(\pi^+\pi^-)\right)\left(1-\frac{\mathcal{A}\left(D^0 \rightarrow K^0(\pi^+\pi^-)\right)}{\mathcal{A}\left(D^0 \rightarrow \bar{K}^0(\pi^+\pi^-)\right)}\right) .
\end{aligned}
\end{equation}
Defining $\mathcal{A}\left(D^0 \rightarrow K^0(\pi^+\pi^-)\right)/\mathcal{A}\left(D^0 \rightarrow \bar{K}^0(\pi^+\pi^-)\right)=-\hat{\rho}_{k_{\mathrm{CP}}} \tan ^2 \theta_C$, the $CP$ transition amplitude for $K_{\text{S}}^0\pi^+\pi^-$ can then be written as
\begin{equation}
\mathcal{A}\left(D^0 \rightarrow K_{\text{S}}^0(\pi\pi)_{k_{\mathrm{CP}}}\right)=\frac{1}{\sqrt{2}} \mathcal{A}\left(D^0 \rightarrow \bar{K}^0(\pi^+\pi^-)\right)\left(1+\hat{\rho}_{k_{\mathrm{CP}}} \tan ^2 \theta_C\right).
\end{equation}
The $U$-spin breaking parameter, $\hat{\rho}~(=|\rho|e^{i (\arg\hat{\rho})})$, is a complex and purely empirical quantity. For a three-body decay with multiple exclusive $CP$ eigenstate resonant contributions $(k_{\mathrm{CP}})$, the $\hat{\rho}$ factors for each are denoted by $\hat{\rho}_{k_{\mathrm{CP}}}$. The $\hat{\rho}$ parameters primarily carry the phase shifts generated as a result of DCS interference, and the magnitudes $(|\hat{\rho}|)$ are naively expected to be of order $ O(1)$. However, the magnitudes are to be empirically measured for the possibility of deviation from the nominal $\tan^2\theta_{C}$ Cabibbo factor. A similar treatment of the decay $D^0 \to K_{\text{L}}^0\pi^+\pi^-$ yields
\begin{equation}
\begin{aligned}
\mathcal{A}\left(D^0 \rightarrow K_{\text{L}}^0(\pi\pi)_{k_{\mathrm{CP}}}\right) & =\frac{1}{\sqrt{2}}\left(\mathcal{A}\left(D^0 \rightarrow \bar{K}^0(\pi^+\pi^-)\right)+\mathcal{A}\left(D^0 \rightarrow K^0(\pi^+\pi^-)\right)\right) \\
& =\frac{1}{\sqrt{2}} \mathcal{A}\left(D^0 \rightarrow \bar{K}^0(\pi^+\pi^-)\right)\left(1+\frac{\mathcal{A}\left(D^0 \rightarrow K^0(\pi^+\pi^-)\right)}{\mathcal{A}\left(D^0 \rightarrow \bar{K}^0(\pi^+\pi^-)\right)}\right),
\end{aligned}
\end{equation}
and
\begin{equation}
\mathcal{A}\left(D^0 \rightarrow K_{\text{L}}^0(\pi\pi)_{k_{\mathrm{CP}}}\right)=\frac{1}{\sqrt{2}} \mathcal{A}\left(D^0 \rightarrow \bar{K}^0(\pi^+\pi^-)\right)\left(1-\hat{\rho}_{k_{\mathrm{CP}}} \tan ^2 \theta_C\right) .
\end{equation}
It is then straightforward to show that the $CP$ resonant amplitudes of the $D^0 \to K_{\text{L}}^0\pi^+\pi^-$ decay can be related to the $D^0 \to K_{\text{S}}^0(\pi^+\pi^-)_{k_{\mathrm{CP}}}$ amplitudes as
\begin{equation}
\begin{aligned}
&\frac{\mathcal{A}\left(D^0 \rightarrow K_{\text{L}}^0(\pi\pi)_{k_{\mathrm{CP}}}\right)}{\mathcal{A}\left(D^0 \rightarrow K_{\text{S}}^0(\pi\pi)_{k_{\mathrm{CP}}}\right)}=\frac{1-\hat{\rho}_{k_{\mathrm{CP}}} \tan ^2 \theta_C}{1+\hat{\rho}_{k_{\mathrm{CP}}} \tan ^2 \theta_C} \approx 1-2 \hat{\rho}_{k_{\mathrm{CP}}} \tan ^2 \theta_C+O\left(\hat{\rho}_{k_{\mathrm{CP}}}^2\tan ^4 \theta_C\right) \Rightarrow\\
&\mathcal{A}\left(D^0 \rightarrow K_{\text{L}}^0(\pi\pi)_{k_{\mathrm{CP}}}\right)=\left(1-2 \hat{\rho}_{k_{\mathrm{CP}}} \tan ^2 \theta_C\right) \times \mathcal{A}\left(D^0 \rightarrow K_{\text{S}}^0(\pi\pi)_{k_{\mathrm{CP}}}\right).
\end{aligned}
\label{KLpipi_amp}
\end{equation}
Even for the largest fitted value of $|\rho_{k_{CP}}|^2$, including the higher-order terms does not lead to a significant change in the results.  Therefore, these contributions can be neglected. As briefly mentioned in Sec.~\ref{Intro}, an amplitude model description of the $D^0 \to K_{\text{L}}^0\pi^+\pi^-$ decay mode is required for a constrained strong-phase measurement. The $D^0 \to K_{\text{L}}^0\pi^+\pi^-$ amplitude model can be obtained via simple DCS interference-motivated modifications, such as the one stated in Eq.~\ref{KLpipi_amp}, to the well-studied $D^0 \to K_{\text{S}}^0\pi^+\pi^-$ amplitude model. Another difference is expected in the DCS resonant model such as $D^0 \to (K_{\text{S},\text{L}}^{0}\pi^+)_{K^*}\pi^-$, with a relative minus sign between the $D^0 \to K_{\text{S}}^0\pi^+\pi^-$ and $D^0 \to K_{\text{L}}^0\pi^+\pi^-$ amplitudes, again because of the phase structure in the definition of $D^0 \to K_{\text{S},\text{L}}^0\pi^+\pi^-$ in terms of the flavour states. We insert this minus sign into the DCS amplitudes of $D^0 \to K_{\text{L}}^0\pi^+\pi^-$ instead of $D^0 \to K_{\text{S}}^0\pi^+\pi^-$ to maintain consistency with the standalone $D^0 \to K_{\text{L}}^0\pi^+\pi^-$ amplitude structure with no minus sign in the DCS parts. Doing so merely introduces an extra $180^{\circ}$ phase added to the nominal DCS phases and does not affect physical observations. The total amplitudes are summarised below:
\begin{equation}
\mathcal{A}\left(D^0 \rightarrow K_{\text{S}}^0\pi^+\pi^-\right)=\sum_r \mathcal{A}_{\bar{K}^0\pi\pi}^{\mathrm{CF}}+\sum_{r \prime} \mathcal{A}_{K^0\pi\pi}^{\mathrm{DCS}}+\sum_{k_{\mathrm{CP}}} \mathcal{A}_{K_{\text{S}}^0(\pi^+\pi^-)_{k_{\mathrm{CP}}}}^{\mathrm{CP}},
\label{Amp_KSpipi}
\end{equation}
\begin{equation}
\mathcal{A}\left(D^0 \rightarrow K_{\text{L}}^0\pi^+\pi^-\right)=\sum_r \mathcal{A}_{\bar{K}^0\pi\pi}^{\mathrm{CF}}-\sum_{r \prime} \mathcal{A}_{K^0\pi\pi}^{\mathrm{DCS}}+\sum_{k_{\mathrm{CP}}}\left(1-2 \hat{\rho}_{k_{\mathrm{CP}}} \tan ^2 \theta_C\right) \mathcal{A}_{K_{\text{S}}^0(\pi^+\pi^-)_{k_{\mathrm{CP}}}}^{\mathrm{CP}}.
\label{Amp_KLpipi}
\end{equation}
where the index $r$ runs over all intermediate resonances in the CF decay chain $D^0 \to \bar{K}^0\pi^+\pi^-$ described by $\mathcal{A}_{\bar{K}^0\pi\pi}^{\mathrm{CF}}$, $r'$ runs over those in the DCS chain $D^0 \to K^0\pi^+\pi^-$ described by $\mathcal{A}_{K^0\pi\pi}^{\mathrm{DCS}}$, and $k_{\mathrm{CP}}$ enumerates the different $CP$-eigenstate contributions in the decay $K_{\text{S}}^0(\pi^+\pi^-)_{k_{\mathrm{CP}}}$. The only way to determine the $\hat{\rho}$ parameters associated with each of the two-body intermediate resonant structure contributions in the $D^0 \to K_{\text{S},\text{L}}^0\pi^+\pi^-$ decay is to fit an amplitude model for $D^0 \to K_{\text{L}}^0\pi^+\pi^-$, where the $D^0 \to K_{\text{S}}^0\pi^+\pi^-$ amplitude may be used as a constraint in a simultaneous fit.

We correct for the fact that hadronic decays of the $D^0 \to K^- X$, where $X$ are combinations of pions on the tag side, are not an exact flavour tag. This effect comes from a contamination by tag-side DCS decays of $D^0 \to K^+X, \bar{D}^0 \to K_{\text{S},\text{L}}^0 \pi^+\pi^-$. An event tagged by a DCS decay will incorrectly place the signal $D^0 \to K^+X, \bar{D}^0 \to K_{\text{S},\text{L}}^0 \pi^+\pi^-$ at $(M^2_{K_{\text{S},\text{L}}^0\pi^-},M^2_{K_{\text{S},\text{L}}^0\pi^+})$ rather than $(M^2_{K_{\text{S},\text{L}}^0\pi^+},M^2_{K_{\text{S},\text{L}}^0\pi^-})$. These DCS-tagged events can be accounted for by adding their amplitudes to the nominal CF pure-flavour-tag amplitude coherently as
\begin{equation}
\mathcal{A}=\mathcal{A}_F \bar{\mathcal{A}}_{\bar{G}}-\bar{\mathcal{A}}_{\bar{F}} \mathcal{A}_{\bar{G}},
\end{equation}
where $\mathcal{A}_{F}(\bar{\mathcal{A}}_{\bar{F}})$ is the amplitude of the signal decay $D^0(\bar{D}^0)\to K_{\text{S},\text{L}}^0\pi^+\pi^-$ and $\bar{\mathcal{A}}_{\bar{G}}(\mathcal{A}_{\bar{G}})$ is the amplitude of the tag decay $\bar{D}^0~(D^0) \to K^+\pi^-$. The decay probabilities, integrated over the allowed tag-mode PHSP, at position $\textbf{x}$ on the $D^0 \to K_\text{S}^0\pi^+\pi^-$ and $D^0 \to K_{\text{L}}^0\pi^+\pi^-$ DPs, are given by
\begin{equation}
|\mathcal{A}(\mathbf{x})|_{K_{\text{S}}^0 \pi \pi}^2=\left(\left|\mathcal{A}_F(\mathbf{x})\right|^2+\left(r_D^G\right)^2\left|\bar{\mathcal{A}}_{\bar{F}}(\mathbf{x})\right|^2-2 r_D^G R_G \mathrm{Re} \left[e^{-i \delta_G} \mathcal{A}_F^*(\mathbf{x}) \bar{\mathcal{A}}_{\bar{F}}(\mathbf{x})\right]\right),
\label{total_ampKS}
\end{equation}
\begin{equation}
|\mathcal{A}(\mathbf{x})|_{K_{\text{L}}^0 \pi \pi}^2=\left(\left|\mathcal{A}_F(\mathbf{x})\right|^2+\left(r_D^G\right)^2\left|\bar{\mathcal{A}}_{\bar{F}}(\mathbf{x})\right|^2+2 r_D^G R_G \mathrm{Re}\left[e^{-i \delta_G} \mathcal{A}_F^*(\mathbf{x}) \bar{\mathcal{A}}_{\bar{F}}(\mathbf{x})\right]\right),
\label{total_ampKL}
\end{equation}
where the parameter $r_{D}^{G}$ is the DCS to CF amplitude ratio and $R_{G}e^{-i\delta_{G}}$ measures the coherence between them for the tag mode $G$. The last terms in Eqs.~(\ref{Amp_KSpipi}) and (\ref{Amp_KLpipi}) contain the interference between CF and DCS amplitudes, which dominates the tag-side DCS effects in the total decay probability. The values used for the hadronic parameters for the three tag modes involved in this analysis are from Refs.~\cite{DCS_1} and \cite{DCS_2} and are listed in Table~\ref{table_DCS}.

\begin{table}[htbp]
\centering
\caption{\small The values of \( r, R, \delta \) for various tag modes.}
\label{table_DCS}
\begin{tabular}{c|c|c|c}
\hline \hline
\text{Tag mode} & \( r^G_D\,(\%) \) & \( R_G \) & \( \delta_G\,(^\circ) \) \\
\hline
\( K^{+} \pi^{-} \) & \( 5.855 ^{+ 0.010}_{- 0.009} \)~\cite{DCS_1} & 1 & \( 191.6_{-2.4}^{+2.5} \)~\cite{DCS_1} \\
\( K^{+} \pi^{-} \pi^0 \) & \( 4.41 \pm 0.11 \)~\cite{DCS_2} & \( 0.79 \pm 0.04 \)~\cite{DCS_2} & \( 196 \pm 11 \)~\cite{DCS_2} \\
\( K^{+} \pi^{-} \pi^{-} \pi^{+} \) & \( 5.50 \pm 0.07 \)~\cite{DCS_2} & \( 0.44_{-0.10}^{+0.09} \)~\cite{DCS_2} & \( 161_{-18}^{+28} \)~\cite{DCS_2} \\
\hline \hline
\end{tabular}
\end{table}

The signal PDF $f_{\text{sig}}(\textbf{x})$ at a PHSP vector $\textbf{x}$ is expressed in terms of the total amplitude $\mathcal{A}_{\text{sig}}(\textbf{x})$ obtained from the isobar model and the detector efficiency $\epsilon(\textbf{x})$ as
\begin{equation}
f_{\text{sig}}(\textbf{x}) = \frac{\epsilon(\textbf{x})|\mathcal{M}(\textbf{x})|^2}{\int \epsilon(\textbf{x})|\mathcal{M}(\textbf{x})|^2\mathrm{d}\textbf{x}}.
\label{signal-PDF}
\end{equation}

The integral can be approximated as a discrete summation of the signal PDF over a large number $N$ of PHSP points, called the integration sample:
\begin{equation}
  \int \epsilon(\textbf{x})|\mathcal{M}(\textbf{x})|^2\mathrm{d}\textbf{x} \propto \frac{1}{N_{\text{MC}}}\sum^{N_{\text{MC}}}_{k_{\text{MC}}}\frac{|\mathcal{M}(\textbf{x})|^2}{|\mathcal{M}^{\text{gen}}(\textbf{x})|^2},
  \label{integral}
\end{equation}
where $k_{\text{MC}}$ is the index of the $k^{\text{th}}_{\text{MC}}$ event of the MC samples and
$N_{\text{MC}}$ is the number of the selected MC sample events. The function $\mathcal{M}^{\text{gen}}(\textbf{x})$ denotes the probability density used to generate the MC sample in 
the MC integration. Since the sample is generated uniformly over the phase space, $\mathcal{M}^{\text{gen}}(\textbf{x})$ is constant. In this analysis, a PHSP MC sample with about 100 times the data size is used in the normalisation integral calculation.

The background PDF is given by 
\begin{equation}
  f_B(\textbf{x})=\frac{B(\textbf{x})}{\int{B(\textbf{x})}\mathrm{d}\textbf{x}},	
  \label{pdf:bkg}
\end{equation}
where $B(\textbf{x})$ represents the background function.  To extract the shared component $\epsilon(\textbf{x})$, the background PDF can be expressed as
\begin{equation}
  f_B(\text{x})=\frac{\epsilon(\textbf{x})B_{\epsilon}(\textbf{x})}{\int{\epsilon(\textbf{x})B_{\epsilon}(\textbf{x})}\mathrm{d}\textbf{x}},	
  \label{pdf:bkg_new}
\end{equation}
where $B_{\epsilon}(\textbf{x}) = B(\textbf{x}) / \epsilon(\textbf{x})$ is the efficiency-corrected background function. The background function in the data is modelled by the background events in the signal region derived from the inclusive MC sample. The invariant-mass distributions of events outside the signal region show good agreement between the data and MC simulation, thus validating the description from the inclusive MC sample. 

The invariant-mass squared distributions $M^2_{K^0_{\text{S},\text{L}}\pi^-},M^2_{K^0_{\text{S},\text{L}}\pi^+},M^2_{\pi^+\pi^-}$ are modelled with the XGBoost package~\cite{XG1,XG2}. By training the inclusive MC samples and PHSP MC samples using different XGBoost models, we obtain the background probability $B(\text{x})R_3(\text{x})$ and the efficiency probability $\epsilon(\text{x})R_3(\text{x})$, respectively. Dividing the two probabilities yields the value of $B_{\epsilon}(\text{x})$.

By combining Eq.~\ref{signal-PDF} and Eq.~\ref{pdf:bkg_new}, the $\epsilon(\textbf{x})$ term is independent of the fitted variables; they are regarded as constants in the fit. As a consequence, the log-likelihood becomes
\begin{equation}
\ln\mathcal{L}=\sum^{2}_{i=1}\sum^{3}_{j=1}\sum\limits_{k=1}^{N^{\text{data}}_{ij}}\ln\bigg[\omega_{\text{sig},ij}\frac{|\mathcal{M}_{ij}(\textbf{x}_{ijk})|^2}{\int \epsilon_{ij}(\textbf{x})|\mathcal{M}_{ij}(\textbf{x})|^2 \mathrm{d}\textbf{x}}+(1-\omega_{\text{sig},ij})\frac{B_{\epsilon,ij}(\textbf{x}_{ijk})}{\int{\epsilon_{ij}(\textbf{x})B_{\epsilon,ij}(\textbf{x})}\mathrm{d}\textbf{x}}\bigg].
  \label{likelihood}
\end{equation}
where $k$ indicates the $k^{\text{th}}$ event in the $i^{\text{th}}$ decay mode ($D^0 \to K_{\text{S},\text{L}}^0\pi^+\pi^-$) and the $j^{\text{th}}$ ST mode, $N^{\text{data}}$ is the number of surviving events for the $j^{\text{th}}$ ST mode.

The normalisation integrals of the signal and background are evaluated using PHSP MC samples, taking into account the differences in $K^0_\text{S}$, $\pi^-$, and $\pi^+$ reconstruction between data and MC simulation,

\begin{equation}
  \int \epsilon(\textbf{x})|\mathcal{M}(\textbf{x})|^2 \mathrm{d}\textbf{x} \propto \frac{1}{N_{\text{MC}}}\sum^{N_{\text{MC}}}_{k_{\text{MC}}}\frac{|\mathcal{M}(\textbf{x})|^2}{|\mathcal{M}^{\text{gen}}(\textbf{x})|^2},
  \label{MC-intergral-corrected}
\end{equation}

\begin{equation}
\begin{aligned}
    &\int \epsilon(\textbf{x}) B_{\epsilon}(\textbf{x}) \,\mathrm{d}\textbf{x} \propto \frac{1}{N_{\text{MC}}} \sum_{k_{\text{MC}}}^{N_{\text{MC}}} \frac{ B_{\epsilon}(\textbf{x}^{k_{\text{MC}}}) }{\left|\mathcal{M}^{\text{gen}}(\textbf{x}^{k_{\text{MC}}})\right|^{2}}\,.
\label{bkg-intergral-corrected}
\end{aligned}
\end{equation}
where $k_{\text{MC}}$ is the index of the $k^{\text{th}}$ event of the PHSP MC sample, and $N_{\text{MC}}$ is the number of the selected PHSP MC events. 
The factor $\gamma_{\epsilon}(\textbf{x})$ is calculated as
\begin{equation}
  \gamma_{\epsilon}(\textbf{x}_j) = \prod_{i} \frac{\epsilon_{i,\text{data}}(\textbf{x}_j)}{\epsilon_{i,\text{MC}}(\textbf{x}_j)},
  \label{pwa:gamma}
\end{equation}
where $i$ refers to tracking, PID, and $K_\text{S}^0$ reconstruction; $\epsilon_{i,\text{data}}(\textbf{x}_j)$ and $\epsilon_{i,\text{MC}}(\textbf{x}_j)$ are their efficiencies as a function of the momenta of the daughter particles for data and MC simulation, respectively. 
The normalisation integral is evaluated using PHSP MC samples, which significantly improves the computational efficiency of the MC integration. These samples inherently account for the detection resolution.

\subsubsection{Blatt-Weisskopf barrier factors}\label{Blatte}
The Blatt-Weisskopf barrier factors $X_L(q)$~\cite{Blatte} are the barrier functions for a two-body decay process $a \to bc$. These functions depend on the angular momentum $L$ and the momentum $q$ of the final-state particle $b$ or $c$ in the rest frame of $a$. They are given by
\begin{equation}
\begin{aligned}
 X_{L=0}(q)&=1,\\
 X_{L=1}(q)&=\sqrt{\frac{z_0^2+1}{z^2+1}},\\
 X_{L=2}(q)&=\sqrt{\frac{z_0^4+3z_0^2+9}{z^4+3z^2+9}}\,,
\end{aligned}
\end{equation}
where $z=qR_r$, $z_0=q_0R_r$ and the effective radius, $R_r$, of the barrier is fixed to 1.5~$(\text{GeV}/c)^{-1}$ for the intermediate 
resonances and 5.0~$(\text{GeV}/c)^{-1}$ for the $D^0$ meson. The momentum $q$ is given by
\begin{equation}
q = \sqrt{\frac{(s_a+s_b-s_c)^2}{4s_a}-s_b}\,, \label{q2}
\end{equation}
where $s_a, s_b$ and $s_c$ are the invariant masses squared of particles $a, b$ and $c$, respectively. The value of $q_0$ is that of $q$ when $s_a = m_a^2$, where $m_a$ is the mass of particle $a$.


\subsubsection{Propagator}\label{sec:propagator}
The intermediate resonances $K^*(892)^{\pm}$, $f_{2}(1270)$, $\omega(782)$, $K^{*}(1410)^{\pm}$, $K^{*}(1680)^{-}$, and $K^{*}_{2}(1430)^{\pm}$ are
parameterised with a relativistic Breit-Wigner function,
\begin{equation}
\begin{aligned}
    P(m) &= \frac{1}{m_{0}^{2} - m^2 - im_{0}\Gamma(m)},\\ 
    \Gamma(m) &= \Gamma_{0}\left(\frac{q}{q_{0}}\right)^{2L+1}\left(\frac{m_{0}}{m}\right)X^{2}_{L}(q),
  \label{RBW}
\end{aligned}
\end{equation}
where $m$ is the invariant mass of the decay products, and $m_0$ and $\Gamma_0$ are the mass and width of the intermediate resonance, which are fixed to their known values~\cite{PDG}. The energy-dependent width is denoted by $\Gamma(m)$. $X_L(q)$ is the Blatt-Weisskopf barrier factor, defined in Sec.~\ref{Blatte}.

The $\rho(770)$ and $\rho(1450)$ resonances are parameterised with the Gounaris-Sakurai (GS) line shape~\cite{GS}. The corresponding propagator  is given by
\begin{equation}
P_{\mathrm{GS}}(m)=\frac{1+d\frac{\Gamma_0}{m_0}}{(m^2_0-m^2)+f(m)-im_0\Gamma(m)},
\end{equation}
where
\begin{equation}
	f(m)=\Gamma_0\frac{m^2_0}{q^3_0}\left[q^2(h(m)-h(m_0))+(m^2_0-m^2)q^2_0\frac{dh}{d(m^2)}\Big|_{m^2=m^2_0}\right],
\end{equation}
and the function $h(m)$ is defined as
\begin{equation}
h(m)=\frac{2}{\pi}\frac{q}{m}\ln\left(\frac{m+2q}{2m_{\pi}}\right).
\end{equation}

Its derivative, evaluated at $m=m_0$, is
\begin{equation}
\frac{dh}{d(m^2)}\Big|_{m^2=m_0^2} = h(m_0)[(8q^2_0)^{-1}-(2m^2_0)^{-1}]+(2\pi m^2_0)^{-1},
\end{equation}
where $m_{\pi}$ is the known mass of the $\pi$ meson~\cite{PDG}, and the normalisation condition at $P_{\mathrm{GS}}(0)$ fixes the parameter $d=\frac{f(0)}{\Gamma_0m_0}$. It is found to be
\begin{equation}
d=\frac{3}{\pi}\frac{m^2_{\pi}}{q^2_0}\ln\left(\frac{m_0+2q_0}{2m_{\pi}}\right)+\frac{m_0}{2\pi q_0}-\frac{m^2_{\pi}m_0}{\pi q^3_0}.
\end{equation}

For the $K_{\text{S},\text{L}}^0\pi^{\pm}$ $\text{S}$-wave contribution, we use a parameterisation from scattering data~\cite{kpiswave}, which is described by a $K^*_0(1430)$ Breit-Wigner together with an effective range nonresonant component with a phase shift:
\begin{equation}
A(m) = F\sin\delta_F e^{i\delta_F} + R\sin\delta_R e^{i\delta_R}e^{i2\delta_F},
\end{equation}
with
\begin{equation}
\begin{aligned}
&\delta_F = \phi_F + \cot^{-1}\Big[\frac{1}{aq}+\frac{rq}{2}\Big],\\
&\delta_R = \phi_R + \tan^{-1}\Big[\frac{M\Gamma(m_{K\pi})}{M^2-m^2_{K\pi}}\Big],
\end{aligned}
\end{equation}
where the parameters $F(\phi_{F})$ and $R(\phi_{R})$ are the magnitude (phase) for the non-resonant state and the resonance terms, respectively, and $a$ and $r$ are the scattering length and the effective interaction length. $q$ and $\Gamma(m_{K\pi})$ are defined in the same way as above. The parameters values ($M$, $\Gamma$, $\phi_{F}$, $a$, $r$) are taken from Ref.~\cite{kpiswave} and fixed in the present fit. They are summarized in Table~\ref{stable} together with the fit results.
\begin{table}[hbtp]
    \centering
    \small
    \caption{$K\pi$ $\text{S}$-wave parameters, obtained from the fit to the $D^0 \to K_{\text{S}}^0 \pi^{+} \pi^{-}$ DP distribution by BABAR and Belle.}
    \begin{tabular}{l c}
    \hline
    \hline
      $M$ ($\mathrm{GeV}/c^{2}$) & $1.441 \pm 0.002$  \\
      $\Gamma$ $(\mathrm{GeV})$ & $0.193 \pm 0.004$ \\
      $F$ & $0.96 \pm 0.07$ \\
      $\phi_{F}$ $(^\circ)$ & $0.1 \pm 0.3$ \\
      $R$ & 1~(fixed) \\
      $\phi_{R}$ $(^\circ)$ & $-109.7 \pm 2.6$ \\
      $a$ $(\text{GeV}/c)^{-1}$ & $0.113 \pm 0.006$ \\
      $r$ $(\text{GeV}/c)^{-1}$ & $-33.8 \pm 1.8$ \\
  \hline
  \hline
    \end{tabular}
    \label{stable}
\end{table}

For the $\pi^+\pi^-$ $\text{S}$-wave component, we use the $K$-matrix parameterisation to describe it. A detailed description of the $K$-matrix formalism can be found in various Refs.~\cite{kpiswave,pipiswave}.
We refer to the ``$K$-matrix amplitude'' as a product of the vector $P$ and the matrix propagator $(I-iK\rho)^{-1}$:
\begin{equation}
    A_{i} = (I-iK\rho)^{-1}_{ij}P_{j},
\end{equation}
where $I$ is the identity matrix, $K$ is the $K$-matrix describing the scattering process and $\rho$ is the PHSP matrix. The indices $i$ and $j$ represent the coupled channels ($1=\pi\pi,2=K\bar{K},3=4\pi,4=\eta\eta,5=\eta\eta'$).

The $K$-matrix is defined as 
\begin{equation}
    K_{ij}(s) = \left(\sum_{\alpha}\frac{g_{i}^{\alpha}g_{j}^{\alpha}}{m_{\alpha}^{2}-s}+f_{ij}^{\mathrm{scatt}}\frac{1-s_{0}^{\mathrm{scatt}}}{s-s_{0}^{\mathrm{scatt}}}\right)f_{A0}(s),
\end{equation}
where $g_{i}^{\alpha}$ denote the real coupling constants of the pole $m_{\alpha}$ to meson channel $i$. The parameters $f_{ij}^{\mathrm{scatt}}$ and $s_{0}^{\mathrm{scatt}}$ describe a smooth part of the $K$-matrix elements. The number 1 is understood to carry units of $\text{GeV}^{2}/c^{4}$. $s$ is the square of the invariant mass of the $\pi^+\pi^-$ system. The symbol $f_{A0}$ is the Adler zero factor, defined as
\begin{equation}
    f_{A0}=\frac{1-s_{A_{0}}}{s-{s_{A_{0}}}}(s-s_{A}m_{\pi}^{2}/2).
\end{equation}

The Adler zero factor suppresses the false kinematic singularity at $s=0$ in the physical region close to the $\pi^+\pi^-$ threshold~\cite{PhysRev.137.B1022}. The units of $s_{0}^{\text{scatt}}$ and $s_{A0}$ are $\text{GeV}^{2}/c^{4}$.

The $P$ vector is given by
\begin{equation}
    P_{j}(s)= f_{1j}^{\mathrm{prod}}\frac{1-s_{0}^{\mathrm{prod}}}{s-s_{0}^{\mathrm{prod}}}+\sum_{\alpha}\frac{\beta_{\alpha}g_{j}^{\alpha}}{m_{\alpha}^{2}-s},
\end{equation}
where the $\beta_{\alpha}$ are the complex production couplings, and $f_{1j}^{\mathrm{prod}}$ and $s_0^{\mathrm{prod}}$ describe the production of the slowly-varying part of the $K$-matrix. In this analysis, the $K$-matrix parameters $m_{\alpha}$, $g_i^{\alpha}$, $f_{ij}^{\mathrm{scatt}}$, $s_0^{\mathrm{scatt}}$, $s_{A0}$, $s_A$, $f_{13}^{\mathrm{prod}}$, $f_{14}^{\mathrm{prod}}$, $f_{15}^{\mathrm{prod}}$, $\beta_{3}$, $\beta_4$, $\beta_5$, and $s_0^{\mathrm{prod}}$ are fixed to the results in Ref.~\cite{kpiswave}. The parameters $f_{11}^{\mathrm{prod}}$, $f_{12}^{\mathrm{prod}}$, $\beta_{1}$, and $\beta_{2}$ are obtained from our fit.

\subsubsection{Spin factors}\label{sec:spinfactor}
The spin factor $S_{i}$ for the process $D \to Cr, r\to AB$ is constructed based on the Zemach formalism~\cite{spin}. The amplitudes for resonances with angular momenta larger than two are not considered due to the limited phase space. The spin factor is expressed as
\begin{equation}
\begin{aligned}
&S_{(0)}   = 1, &(S~\text{wave})\\
&S_{(1)} =M^{2}_{BC}-M_{AC}^{2}+\frac{(M_{D}^{2}-M_{C}^{2})(M_{A}^{2}-M_{B}^{2})}{M_{AB}^{2}}, \qquad &(P~\text{wave})\\
&S_{(2)}=a_{1}-\frac{1}{3}a_{2}a_{3}. &(D~\text{wave})
\end{aligned}
\end{equation}
where $a_1, a_2$ and $a_3$ are given by
\begin{equation}
\begin{aligned}
&a_1 = \left[M_{BC}^{2} - M_{AC}^{2} + \frac{(M_{D}^{2}-M_{C}^{2})(M_{A}^{2}-M_{B}^{2})}{M_{AB}^{2}} \right]^{2},\\
&a_{2} = M_{AB}^{2} - 2M_{D}^{2} - 2M_{C}^{2} + \frac{(M_{D}^{2}-M_{C}^{2})^{2}}{M_{AB}^{2}} ,\\
&a_{3} =M_{AB}^{2}-2M_{A}^{2}-2M_{B}^{2}+\frac{(M_{A}^{2}-M_{B}^{2})^{2}}{M_{AB}^{2}}.
\end{aligned}
\end{equation}
Here, $M_{AB}, M_{AC}$, and $M_{BC}$ denote the invariant masses of the particle combinations $AB, AC$, and $BC$, respectively, and $M_{A}, M_{B}, M_{C}$ and $M_{D}$ denote the nominal masses of $A, B, C$, and $D$, respectively.


\subsection{Fit results}\label{sec:fitresults}
\begin{figure*}[htbp] 
  \centering
  \includegraphics[width=4.9cm]{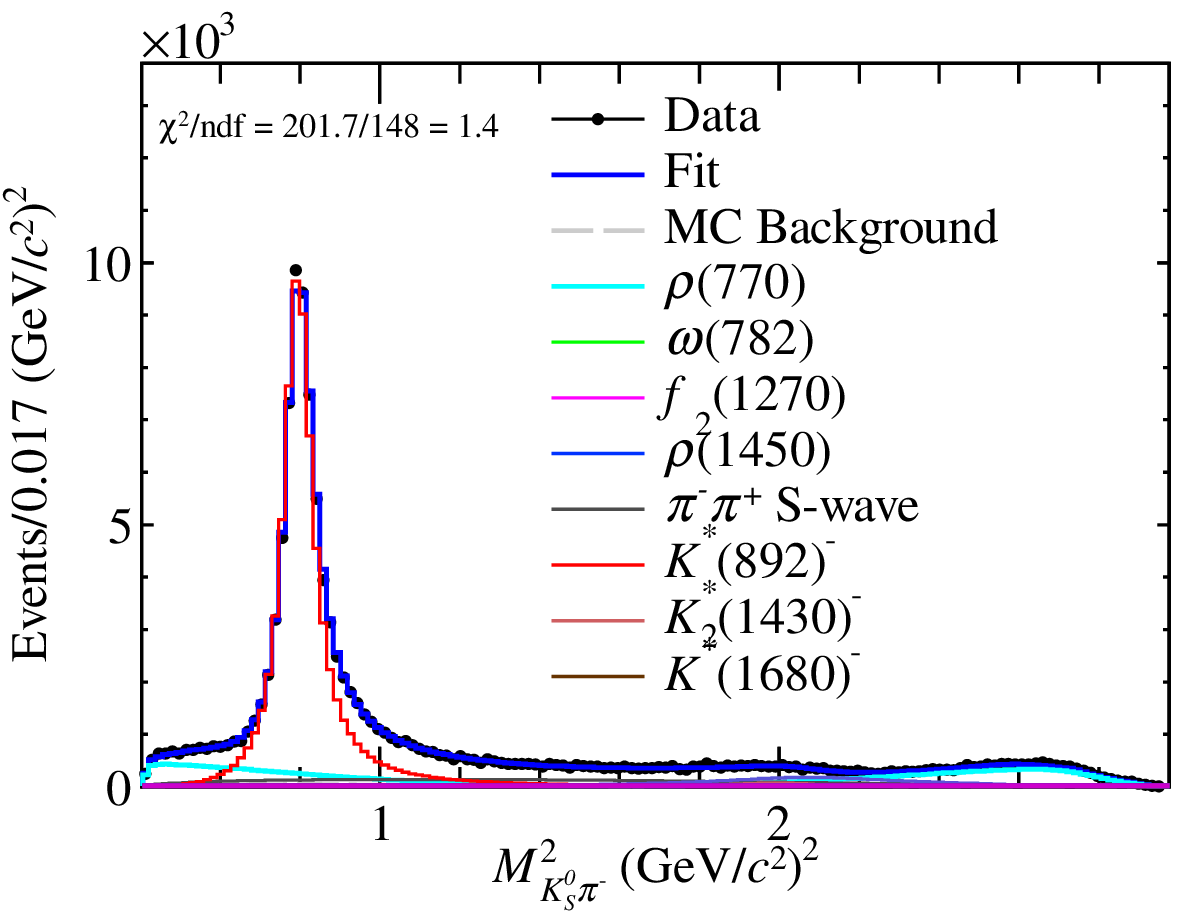}
  \includegraphics[width=4.9cm]{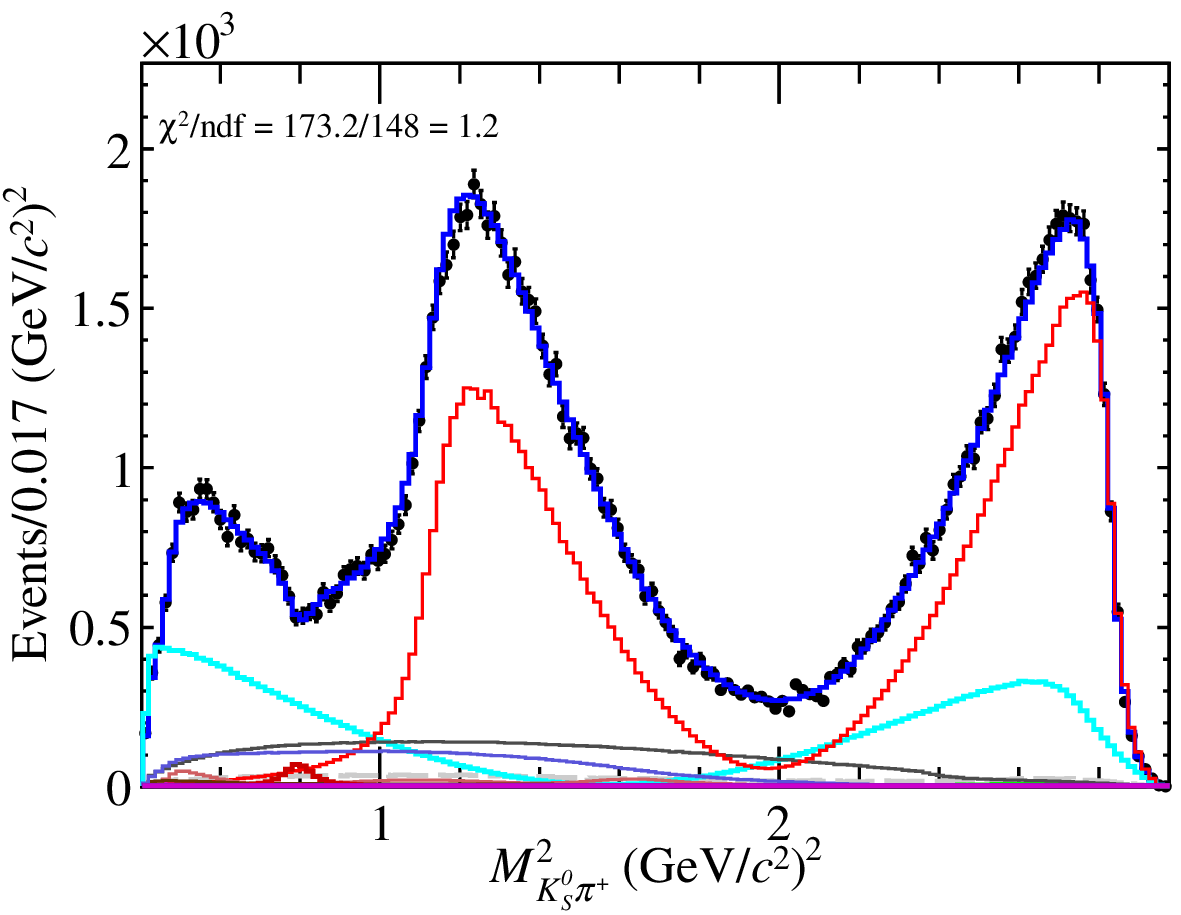}
  \includegraphics[width=4.9cm]{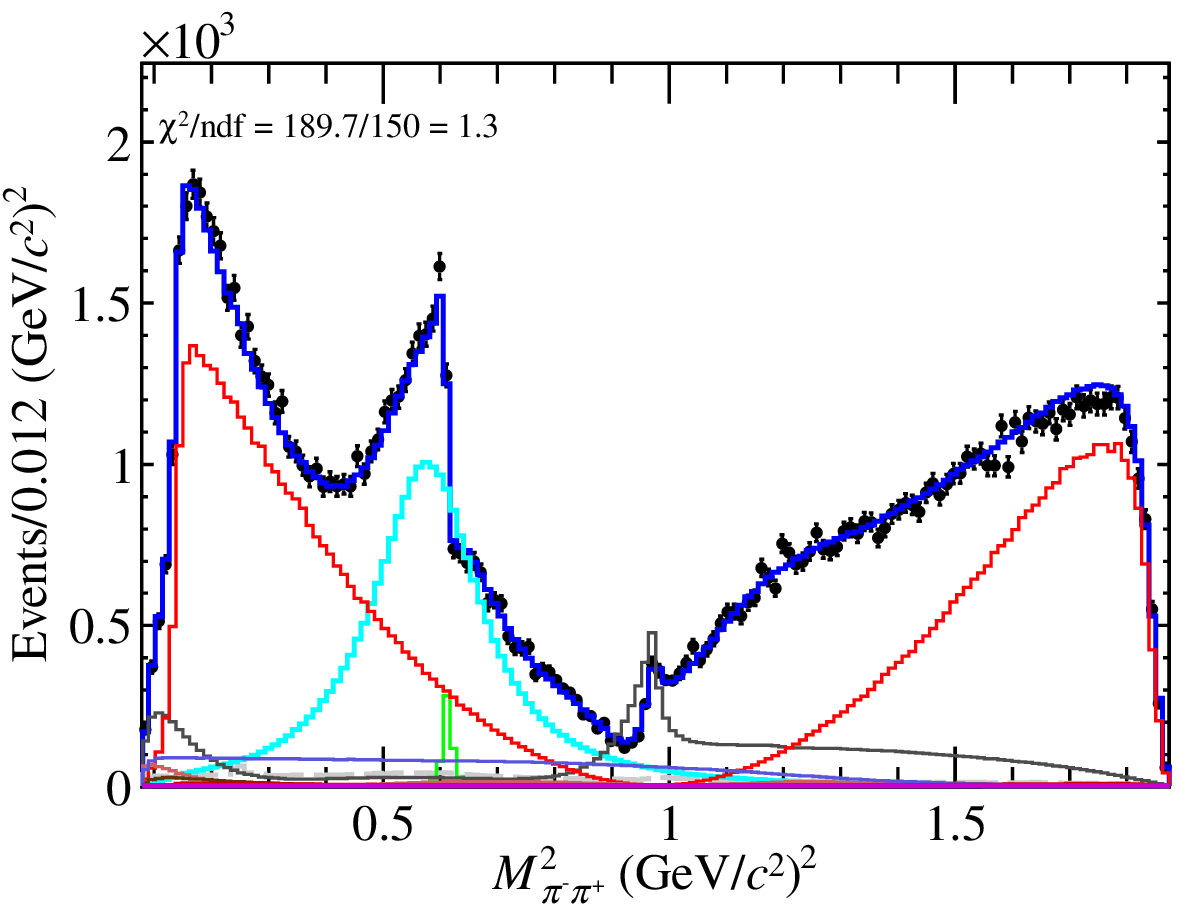}\\
  \includegraphics[width=4.9cm]{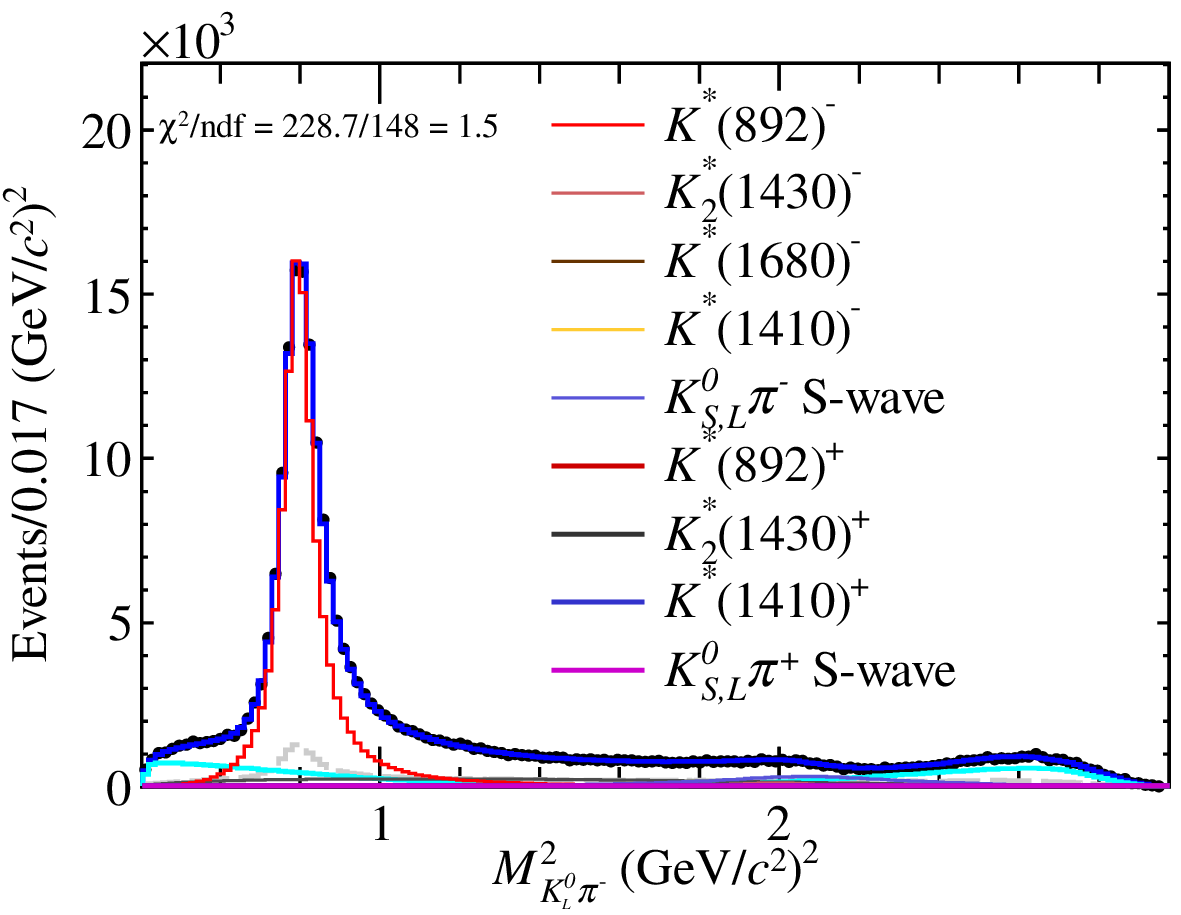}
  \includegraphics[width=4.9cm]{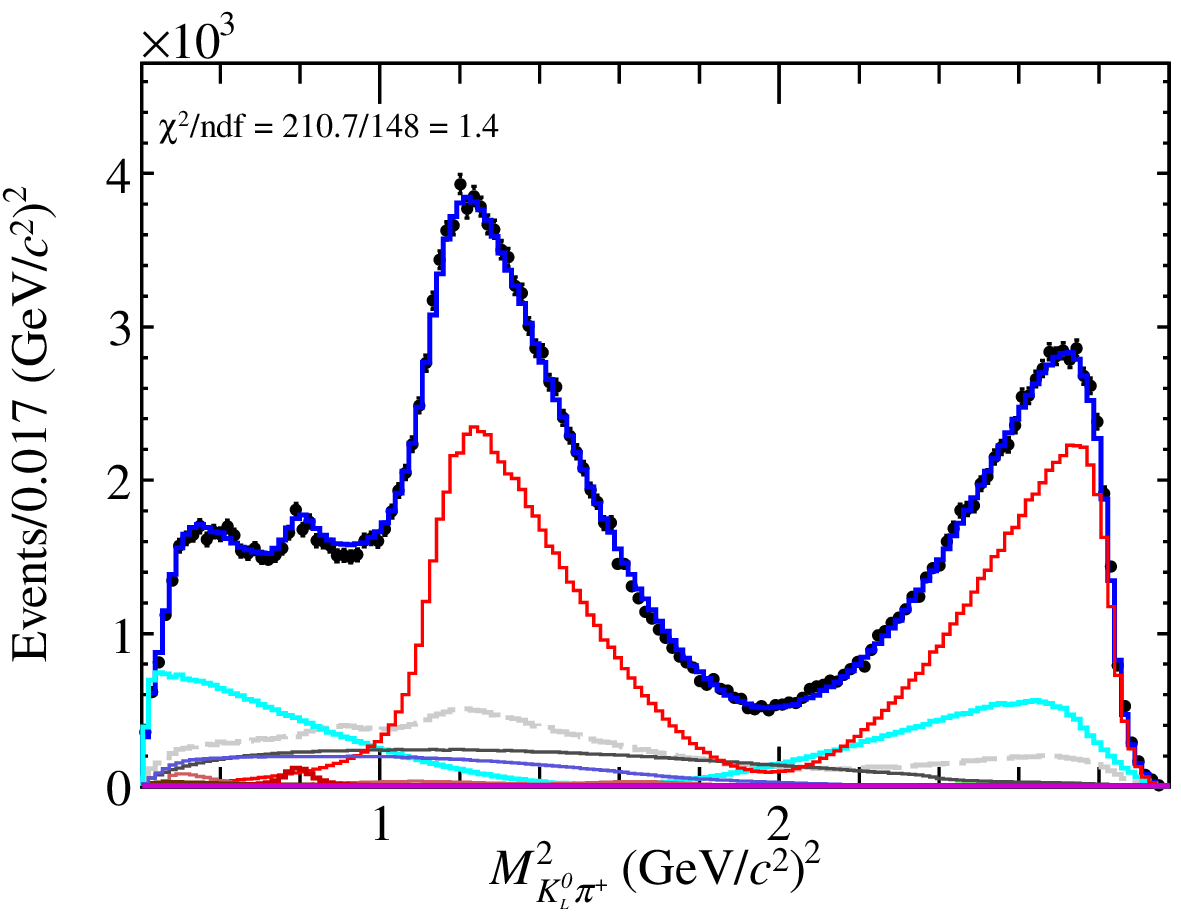}
  \includegraphics[width=4.9cm]{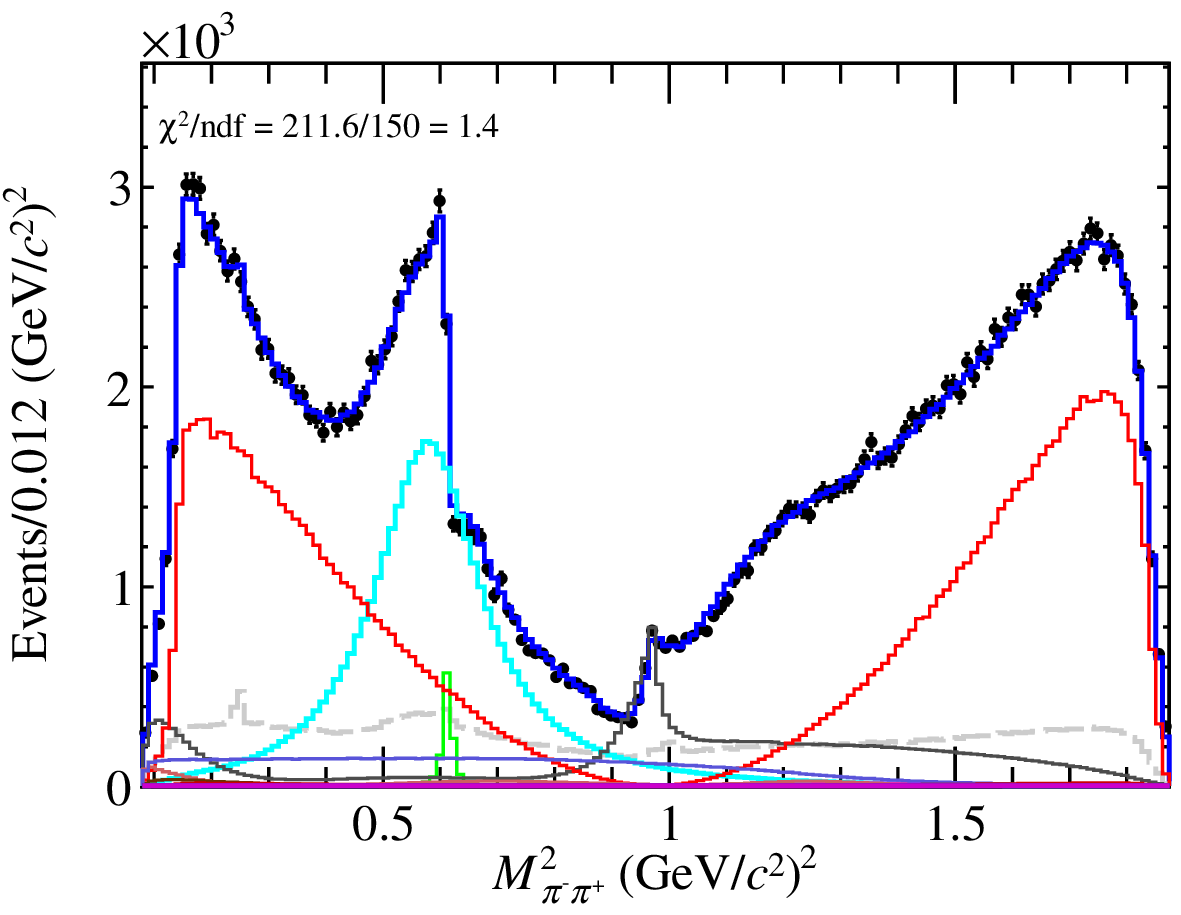}\\
  \caption{Distributions of $M^{2}_{K_{\text{S},\text{L}}^0\pi^-}$, $M^{2}_{K_{\text{S},\text{L}}^0\pi^+}$ and $M^{2}_{\pi^+\pi^-}$ from the $D^0 \to K_{\text{S},\text{L}}^{0}\pi^{+}\pi^{-}$ simultaneous fit. The points with error bars are data, and the blue lines are the total fit result. The dashed grey lines are the simulated background derived from the inclusive MC sample.}
  \label{fig:KSLpipi_amp_plot}
\end{figure*}

In the fit, the magnitude and phase of the process $D^{0} \to K_{\text{S},\text{L}}^{0}\rho(770)$ are fixed to be 1.0 and 0.0. All the possible combinations with different intermediate processes are tested, and only those processes with a statistical significance larger than $5\sigma$ are retained, where the statistical significance of each amplitude is calculated based on the change in the log-likelihood with and without this amplitude after taking the change in the number of degrees of freedom into account. The final model contains fourteen components. The projections of the invariant mass distributions are shown in Fig.~\ref{fig:KSLpipi_amp_plot}. The magnitudes, phases, fit fractions (FFs), and $U$-spin breaking parameters are listed in Table~\ref{tab:amp_result} and Table~\ref{tab:uspin}. The FF of the $n^{\text{th}}$ amplitude is calculated by
\begin{equation}
\mathrm{FF}_n=\int\left|a_n e^{i \phi_n} \mathcal{A}_n\right|^2 \mathrm{d}\Phi_3 / \int|\mathcal{M}|^2 \mathrm{d}\Phi_3,
\label{Fit-Fraction-Definition}
\end{equation}
where $\mathrm{d}\Phi_3$ is the standard element of the three-body phase space. 

In practice, the FFs are computed numerically using generator-level PHSP MC events. The discrete form of Eq.~\ref{Fit-Fraction-Definition} becomes 
\begin{equation}
\begin{aligned}
  {\text{FF}}_{n} = \frac{\sum^{N_{\text{gen}}} \left|\rho_{n}e^{i\phi_{n}}\mathcal{A}_{n}\right|^{2}}{\sum^{N_{\text{gen}}} \left|\mathcal{M}\right|^{2}}\,, 
  \label{Fit-Fraction-Definition-MC}
\end{aligned}
\end{equation}

A particularly striking feature is the large asymmetry observed between the $K_{\text{S}}^0\pi^+\pi^-$ and  $K_{\text{L}}^0\pi^+\pi^-$ FFs for the $f_2(1270)$ resonance. As shown in Table~\ref{tab:amp_result}, this is the most pronounced asymmetry among all intermediate resonance states. The effect is naturally attributed to the large value of $\mathrm{Re}[\hat{\rho}]$  obtained for $f_2(1270)$ from the simultaneous fit, as listed in Table~\ref{tab:uspin}. This observation provides a direct and consistent link between the $U$-spin breaking parameters.

\begin{table}[hbtp]
\centering
\caption{Phases and FFs for various amplitudes in the nominal fit. The first and second uncertainties on phases and FFs are statistical and systematic, respectively.}
\small
\label{tab:amp_result}

\begin{tabular}{lrrrr}
\hline
\hline
Resonance      & Phase (rad)       & $K_{\text{S}}^0 \pi^{+} \pi^{-}$ FF (\%) & $K_{\text{L}}^0 \pi^{+} \pi^{-}$ FF (\%) \\
\midrule
$K^0_{\text{S},\text{L}}\rho(770)$    & $0.0$               & $19.68 \pm 0.24 \pm 0.26$& $19.50 \pm 0.22 \pm 0.31$ \\
$K^0_{\text{S},\text{L}}\omega(782)$  & $1.87 \pm 0.04 \pm 0.04$ & $0.50  \pm 0.03 \pm 0.03$& $0.58 \pm 0.03 \pm 0.03$  \\
$K^0_{\text{S},\text{L}}f_2(1270)$    & $-0.70\pm 0.03 \pm 0.04$ & $0.94  \pm 0.06 \pm 0.05$& $0.49 \pm 0.04 \pm 0.03$ \\
$K^0_{\text{S},\text{L}}\rho(1450)$   & $1.48 \pm 0.06 \pm 0.12$ & $0.24  \pm 0.05 \pm 0.04$& $0.21 \pm 0.03 \pm 0.04$  \\
$K^0_{\text{S},\text{L}}\pi\pi$ $\text{S}$-wave &           & $11.51 \pm 0.20 \pm 0.24$& $10.47 \pm 0.17 \pm 0.22$      \\
$\beta_{1}$    & $1.27 \pm 0.01 \pm 0.02$ &                 &         \\
$\beta_{2}$    & $0.35 \pm 0.01 \pm 0.04$ &                 &         \\
$f_{11}^{\text{prod}}$ & $-2.24 \pm 0.01 \pm 0.06$ &           &         \\
$f_{12}^{\text{prod}}$ & $-2.57 \pm 0.03 \pm 0.07$&           &         \\
$K^*(892)^{-}\pi^+$ & $5.47 \pm 0.01 \pm 0.02$ & $57.88 \pm 0.23 \pm 0.37$ & $57.71 \pm 0.19 \pm 0.28$  \\
$K_2^*(1430)^{-}\pi^+$& $-0.77 \pm 0.02 \pm 0.03$& $1.28 \pm 0.05 \pm 0.08$ & $1.27 \pm 0.05 \pm 0.08$   \\
$K^*(1680)^{-}\pi^+$& $-5.12 \pm 0.05 \pm 0.10$& $0.65 \pm 0.08 \pm 0.12$ & $0.60 \pm 0.07 \pm 0.11$  \\
$K^0_{\text{S},\text{L}}\pi^-$ $\text{S}$-wave $\pi^+$& $1.65 \pm 0.01 \pm 0.04$& $6.44 \pm 0.12 \pm 0.35$& $6.35 \pm 0.11 \pm 0.35$\\
$K^*(892)^{+}\pi^-$ & $-3.85 \pm 0.03 \pm 0.03$& $0.40 \pm 0.02 \pm 0.03$ & $0.40 \pm 0.02 \pm 0.03$ \\
\hline
\hline
Resonance      & Phase (rad)   & $K_{\text{S}}^0 \pi^{+} \pi^{-}$ FF  ($\times10^{-4}$) & $K_{\text{L}}^0 \pi^{+} \pi^{-}$ FF ($\times10^{-4}$) \\
\hline
$K^*(1410)^{-}\pi^+$& $4.25 \pm 0.15 \pm 0.29$& $1.76 \pm 0.96 \pm 0.67$ & $1.74 \pm 0.95 \pm 0.65$   \\ 
$K_2^*(1430)^{+}\pi^-$& $-2.13 \pm 0.14 \pm 0.10$& $1.12 \pm 0.40 \pm 0.30$ & $1.11 \pm 0.30 \pm 0.29$   \\
$K^*(1410)^{+}\pi^-$& $-1.63 \pm 0.13 \pm 0.09$ & $1.83 \pm 0.50 \pm 0.22$ & $1.82 \pm 0.50 \pm 0.26$   \\
$K^0_{\text{S},\text{L}}\pi^+$ $\text{S}$-wave $\pi^-$& $1.92 \pm 0.07 \pm 0.08$& $4.56 \pm 0.80 \pm 0.38$& $4.50 \pm 0.80 \pm 0.49$\\
\midrule
Total               &                & $99.52\pm0.25\pm0.63$                      & $97.58\pm0.21\pm0.64$                    \\
\hline
\hline
\end{tabular}
\end{table}

\begin{table}[htbp]
    \centering
    \caption{\small Measured $U$-spin breaking parameters.  The first and second uncertainties on phases and FFs are statistical and systematic, respectively.}
    \label{tab:uspin}
    \begin{tabular}{lrrr}
        \hline
        \hline
        \multicolumn{1}{c}{Resonance} & 
        \multicolumn{1}{c}{$|\hat{\rho}|$} & 
        \multicolumn{1}{c}{$\arg (\hat{\rho})\,\left[{\text{rad}}\right]$} & 
        \multicolumn{1}{c}{$\left|1-2 \hat{\rho} \tan ^2 \theta_C\right|^2$} \\
        \hline
        $\rho(770)$   & $0.62 \pm 0.09 \pm 0.06$ & $-2.73 \pm 0.16 \pm 0.27$ & $1.13 \pm 0.02 \pm 0.02$\\
        $\omega(782)$ & $1.56 \pm 0.53 \pm 0.29$ & $-2.78 \pm 0.25 \pm 0.16$ & $1.34 \pm 0.13 \pm 0.07$\\
        $f_2(1270)$   & $3.34 \pm 0.34 \pm 0.19$ & $-0.73 \pm 0.09 \pm 0.05$ & $0.59 \pm 0.05 \pm 0.03$\\
        $\rho(1450)$  & $4.45 \pm 0.83 \pm 0.60$ & $1.30 \pm 0.19 \pm 0.14$ & $0.97 \pm 0.18 \pm 0.13$\\
        $\pi\pi$ $\text{S}$-wave & $0.77 \pm 0.13 \pm 0.17$ & $-1.74 \pm 0.13 \pm 0.11$ & $1.04 \pm 0.02 \pm 0.02$\\
        \hline
        \hline
    \end{tabular}
\end{table}

\section{Systematic uncertainties}
\label{sec:systematics}
The systematic uncertainties in the amplitude analysis arise from six main sources. The detailed results of the systematic uncertainties are summarised in Tables~\ref{tab:sys_uspin_r},~\ref{tab:sys_uspin_phi},~\ref{tab:sys_KSLpha},~\ref{tab:sys_KSFF} and~\ref{tab:sys_KLFF}. The sum in quadrature of the individual uncertainties is considered as the total uncertainty and these are expressed in units of the corresponding statistical uncertainties.
\begin{itemize}
\item[\uppercase\expandafter{\romannumeral1}] 
Amplitude model: \\
The masses and widths of resonances are varied within their $\pm 1\sigma$ boundaries, as tabulated in Ref.~\cite{PDG}, and two dedicated refits are performed to assess the corresponding systematic uncertainties. For each parameter, the systematic uncertainty is determined as the maximum deviation between the results of these boundary fits and the central fit result. The total amplitude model uncertainty is obtained by combining all individual contributions in quadrature.

\item[\uppercase\expandafter{\romannumeral2}]
Effective radius: \\
The systematic uncertainty on the $R_r$ parameters in the Blatt-Weisskopf factors is obtained by repeating the fit with the effective radii of the intermediate states and the $D^0$ meson varied independently by $\pm 0.5~(\text{GeV}/c)^{-1}$ about their corresponding central values.

\item[\uppercase\expandafter{\romannumeral3}] 
Fit bias:\\
To investigate the possible bias introduced by the fit procedure, an ensemble of 150 signal MC samples is generated based on the results of the amplitude analysis. The fit procedure is repeated for each signal MC sample, and the pull distributions of the amplitude fit results are fitted with a Gaussian function. The uncertainty of the fitted mean values is taken as the corresponding systematic uncertainty.

\item[\uppercase\expandafter{\romannumeral4}]  
Background estimation: \\
In the analysis of the impact of background on the amplitude model, two primary sources of influence are identified: background size and background shape. The background size is associated with the signal purity ($\omega_{\text{sig}}$) in Eq.~\ref{likelihood}.
The systematic uncertainty associated with the $\omega_{\text{sig}}$ parameter is assessed by performing two additional fits, where $\omega_{\text{sig}}$ is fixed at its $\pm 1\sigma$ boundaries. For each observable, the corresponding systematic uncertainty is then taken as the maximum deviation between the results of these boundary fits and the nominal fit result. The background shape is related to the background function $B(\textbf{x})$ in Eq.~\ref{pdf:bkg}. To evaluate the uncertainty from background shape, an alternative background sample is used, where the relative fractions of background processes from direct $q\bar{q}$ production are varied according to the statistical uncertainties of their known cross sections~\cite{Luminosity}. The total background-related uncertainty is taken as the square root of the sum in quadrature of the uncertainties from background size and background shape.

\item[\uppercase\expandafter{\romannumeral5}] 
Experimental effects: \\
To estimate the uncertainties associated with $\gamma_{\epsilon}$, as defined in Eq.~\ref{pwa:gamma}, the amplitude model is refitted by varying the reconstruction efficiencies of $K^0_\text{S}$, $\pi^-$, and $\pi^+$ according to their uncertainties. The maximum deviations between the nominal and refitted parameters are then taken as the systematic uncertainties.

\item[\uppercase\expandafter{\romannumeral6}] 
Quantum correlation correction:  \\
The input parameters of the quantum correlation correction are varied within their $\pm 1\sigma$ uncertainties, and the amplitude analysis fit is repeated. The largest difference from the nominal results is taken as the systematic uncertainty.

\end{itemize}

\begin{table}[htbp]
\centering
\caption{\small Systematic uncertainties on the $U$-spin breaking parameters $|\hat{\rho}|$ for each amplitude in units of the corresponding statistical uncertainty. The source are: (I) amplitude model, (II) effective radius, (III) fit bias, (IV) background estimation, (V) experiment effects, (VI) Quantum correlation correction. }
\label{tab:sys_uspin_r}
\begin{tabular}{c|ccccccc}
\hline 
\hline
Resonance  & I & II & III & IV & V & VI &Total\\
\hline
$\rho(770)$        & 0.56 & 0.08 & 0.20 & 0.15 & 0.07 & 0.20 & 0.65\\
$\omega(782)$      & 0.18 & 0.12 & 0.39 & 0.06 & 0.06 & 0.13 & 0.45\\
$f_2(1270)$        & 0.36 & 0.11 & 0.07 & 0.11 & 0.08 & 0.38 & 0.56\\
$\rho(1450)$       & 0.44 & 0.12 & 0.03 & 0.23 & 0.08 & 0.50 & 0.72\\
$\pi\pi$ $\text{S}$-wave    & 0.75 & 0.15 & 0.17 & 0.39 & 0.07 & 1.01 & 1.34\\
\hline
\hline
\end{tabular}
\end{table}

\begin{table}[htbp]
\centering
\caption{\small Systematic uncertainties on the $U$-spin breaking parameters $\arg(\hat{\rho})$ for each amplitude in units of the corresponding statistical uncertainty. The source are: (I) amplitude model, (II) effective radius, (III) fit bias, (IV) background estimation, (V) experiment effects, (VI) Quantum correlation correction.}
\label{tab:sys_uspin_phi}
\begin{tabular}{c|ccccccc}
\hline 
\hline
Resonance & I & II & III & IV & V & VI & Total\\
\hline
$\rho(770)$        & 0.84 & 0.25 & 0.11 & 0.28 & 0.04 & 1.43 & 1.70\\
$\omega(782)$      & 0.33 & 0.13 & 0.28 & 0.22 & 0.03 & 0.37 & 0.62\\
$f_2(1270)$        & 0.46 & 0.22 & 0.05 & 0.12 & 0.05 & 0.16 & 0.55\\
$\rho(1450)$       & 0.43 & 0.46 & 0.12 & 0.10 & 0.04 & 0.29 & 0.71\\
$\pi\pi$ $\text{S}$-wave    & 0.37 & 0.14 & 0.17 & 0.29 & 0.07 & 0.67 & 0.85\\
\hline
\hline
\end{tabular}
\end{table}

\begin{table}[htbp]
\centering
\caption{\small Systematic uncertainties on the $K_{\text{S},\text{L}}^0 \pi^{+} \pi^{-}$ phase for each amplitude in units of the corresponding statistical uncertainty. The source are: (I) amplitude model, (II) effective radius, (III) fit bias, (IV) background estimation, (V) experiment effects, (VI) Quantum correlation correction.}
\label{tab:sys_KSLpha}
\begin{tabular}{c|ccccccc}
\hline
\hline
Resonance & I & II & III & IV & V & VI & Total\\
\hline
$\rho(770)$                    & -- & -- & --  & --  & --  & --  &--\\
$\omega(782)$              & 0.60 & 0.48 & 0.19 & 0.23 & 0.04 & 0.34 & 0.89 \\
$f_2(1270)$                & 0.97 & 0.93 & 0.01 & 0.10 & 0.05 & 0.14 & 1.33\\
$\rho(1450)$               & 1.57 & 1.10 & 0.05 & 0.35 & 0.03 & 0.22 & 1.96\\
$\pi\pi$ $\text{S}$-wave   & -- & --   & --  & --  & --  & --  &--\\
$\beta_{1}$                & 2.25 & 0.43 & 0.07  & 0.39  & 0.04  & 0.14   & 2.33\\
$\beta_{2}$                 & 4.14 & 1.29 & 0.04  & 0.42  & 0.04  & 0.17  & 4.36\\
$f_{11}^{\text{prod}}$         & 6.32 & 0.95 & 0.03  & 0.37  & 0.02  & 0.11  & 6.40\\
$f_{12}^{\text{prod}}$        & 2.12 & 0.76 & 0.08  & 0.58  & 0.01  & 0.31  &2.35\\
$K^*(892)^{-}$              & 1.13 & 1.09 & 0.20 & 0.27 & 0.01 & 0.82 & 1.80\\
$K_2^*(1430)^{-}$          & 1.53 & 0.72 & 0.20 & 0.28 & 0.01 & 0.14  & 1.73\\
$K^*(1680)^{-}$            & 1.52 & 1.17 & 0.02 & 0.58 & 0.04 & 0.12  & 2.01\\
$K^*(1410)^{-}$             & 1.44 & 0.28 & 0.01 & 1.26 & 0.07 & 0.10 & 1.94\\
$K^0_{\text{S},\text{L}}\pi^-$ $\text{S}$-wave  &4.05 &1.41 &0.21 &0.16 & 0.05& 0.44& 4.32\\
$K^*(892)^{+}$           & 0.59 & 0.39 & 0.11 & 0.16 & 0.02 & 0.64 & 0.97\\
$K_2^*(1430)^{+}$            & 0.62 & 0.22 & 0.09 & 0.14 & 0.03 & 0.19 & 0.71\\
$K^*(1410)^{+}$             & 0.43 & 0.36 & 0.13 & 0.08 & 0.01 & 0.44 & 0.73\\
$K^0_{\text{S},\text{L}}\pi^+$ $\text{S}$-wave &0.95 &0.51 &0.07 &0.10 &0.01 & 0.35& 1.14\\
\hline
\hline
\end{tabular}
\end{table}

\begin{table}[htbp]
\centering
\caption{\small Systematic uncertainties on the $K_{\text{S}}^0 \pi^{+} \pi^{-}$ FFs for each amplitude in units of the corresponding statistical uncertainty. The source are: (I) amplitude model, (II) effective radius, (III) fit bias, (IV) background estimation, (V) experiment effects, (VI) Quantum correlation correction.}
\label{tab:sys_KSFF}
\begin{tabular}{c|ccccccc}
\hline
\hline
Resonance & I & II & III & IV & V & VI &Total\\
\hline
$\rho(770)$                & 0.90 & 0.47 & 0.14 & 0.25 & 0.04 & 0.18 & 1.07\\
$\omega(782)$              & 0.38 & 0.76 & 0.07 & 0.08 & 0.01 & 0.21 & 0.88\\
$f_2(1270)$                & 0.65 & 0.39 & 0.05 & 0.19 & 0.01 & 0.33 & 0.85\\
$\rho(1450)$               & 0.67 & 0.10 & 0.18 & 0.05 & 0.04 & 0.23 & 0.74\\
$\pi\pi$ $\text{S}$-wave & 0.94 & 0.66 & 0.03 & 0.28 & 0.03 & 0.10 & 1.19\\
$K^*(892)^{-}$             & 0.95 & 0.69 & 0.06 & 1.05 & 0.01 & 0.17 & 1.59\\
$K_2^*(1430)^{-}$          & 0.86 & 0.25 & 0.32 & 1.36 & 0.05 & 0.11 & 1.66\\
$K^*(1680)^{-}$            & 1.37 & 0.42 & 0.04 & 0.48 & 0.03 & 0.13 & 1.52\\
$K^*(1410)^{-}$            & 0.39 & 0.23 & 0.01 & 0.52 & 0.05 & 0.09 & 0.70\\
$K^0_{\text{S},\text{L}}\pi^-$ $\text{S}$-wave  &2.45 &0.54 &0.19 &1.49 & 0.07& 0.04& 2.93\\
$K^*(892)^{+}$              & 0.47 & 0.32 & 0.06 & 0.05 & 0.03 & 1.19 & 1.32\\
$K_2^*(1430)^{+}$           & 0.42 & 0.06 & 0.18 & 0.26 & 0.04 & 0.55 & 0.76\\
$K^*(1410)^{+}$             & 0.31 & 0.25 & 0.01 & 0.08 & 0.02 & 0.15 & 0.43\\
$K^0_{\text{S},\text{L}}\pi^+$ $\text{S}$-wave  &0.34 &0.05 &0.22 &0.23 &0.01 & 0.07& 0.47\\
\hline
\hline
\end{tabular}
\end{table}

\begin{table}[htbp]
\centering
\caption{\small Systematic uncertainties on the $K_{\text{L}}^0 \pi^{+} \pi^{-}$ fit fractions for each amplitude in units of the corresponding statistical uncertainty. The source are: (I) amplitude model, (II) effective radius, (III) fit bias, (IV) background estimation, (V) experiment effects, (VI) Quantum correlation correction.}
\label{tab:sys_KLFF}
\begin{tabular}{c|ccccccc}
\hline
\hline
Resonance  & I & II & III & IV & V & VI &Total\\
\hline
$\rho(770)$                & 1.13 & 0.67 & 0.20 & 0.48 & 0.03 & 0.27 & 1.41\\
$\omega(782)$              & 0.48 & 0.87 & 0.02 & 0.37 & 0.06 & 0.10 & 1.07\\
$f_2(1270)$                & 0.74 & 0.17 & 0.21 & 0.12 & 0.02 & 0.22 & 0.83\\
$\rho(1450)$               & 0.83 & 0.70 & 0.01 & 0.32 & 0.04 & 0.39 & 1.20\\
$\pi\pi$ $\text{S}$-wave & 0.79 & 0.86 & 0.02 & 0.41 & 0.04 & 0.29 & 1.27\\
$K^*(892)^{-}$             & 1.04 & 0.62 & 0.14 & 0.75 & 0.13 & 0.34 & 1.48\\
$K_2^*(1430)^{-}$          & 0.65 & 0.26 & 0.29 & 1.40 & 0.04 & 0.13 & 1.60\\
$K^*(1680)^{-}$            & 1.36 & 0.41 & 0.04 & 0.46 & 0.03 & 0.09 & 1.50\\
$K^*(1410)^{-}$            & 0.40 & 0.22 & 0.01 & 0.51 & 0.01 & 0.10 & 0.68\\
$K^0_{\text{S},\text{L}}\pi^-$ $\text{S}$-wave &2.44 &0.52 &0.61 &1.80 & 0.04& 0.41 &3.16\\
$K^*(892)^{+}$             & 0.58 & 0.31 & 0.05 & 0.42 & 0.01 & 1.31 & 1.53\\
$K_2^*(1430)^{+}$          & 0.65 & 0.37 & 0.19 & 0.27 & 0.04 & 0.53 & 0.98\\
$K^*(1410)^{+}$            & 0.42 & 0.19 & 0.01 & 0.15 & 0.02 & 0.18 & 0.52\\
$K^0_{\text{S},\text{L}}\pi^+$ $\text{S}$-wave &0.34 &0.25 &0.21 &0.23 &0.01 & 0.32&0.61\\
\hline
\hline
\end{tabular}
\end{table}

\clearpage

\section{Model-predicted strong-phase parameters}
\begin{figure}[htbp]
    \centering
    \includegraphics[width=0.5\linewidth]{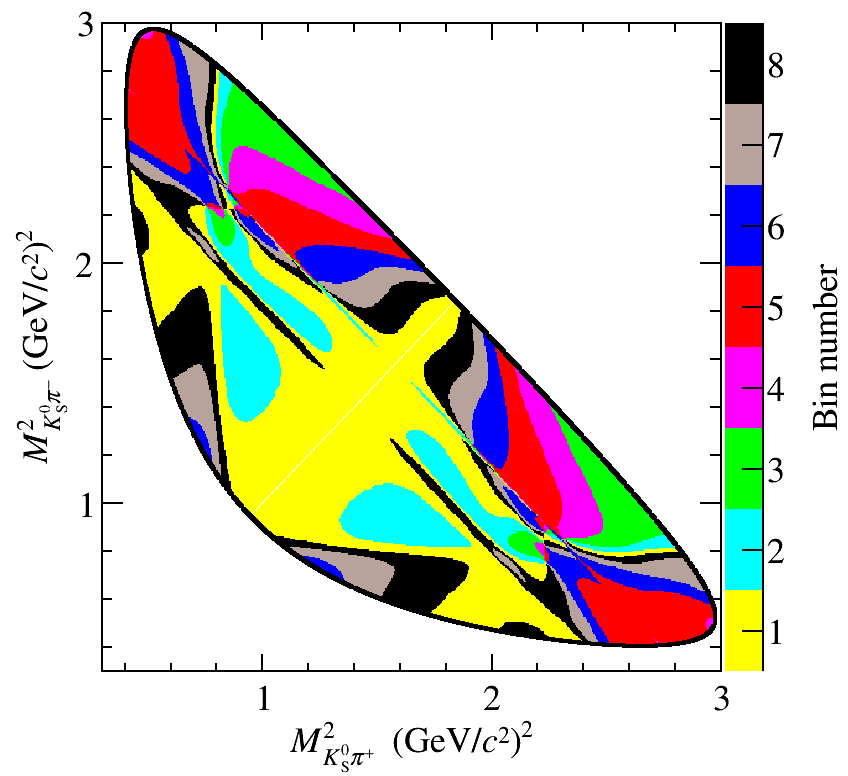}
    \caption{Equal-$\Delta\delta_D$ binning of the DP for $D^0\to K_{\text{S}}^0\pi^+\pi^-$ decays, based on the model in Ref.~\cite{kspipi_8bins}. The colour scale indicates the absolute value of the bin number $i$.}
    \label{fig:bins_KSpipi}
\end{figure}

The $D$-decay parameters that appear in these studies~\cite{cisi_1,cisi_2,cisi_3} are cosines and sines of the strong-phase difference averaged over regions of phase space. The 2D phase space of $K^0_{\text{S}}\pi^+\pi^-$ and $K^0_\text{L}\pi^+\pi^-$ decay modes is described in terms of pairwise invariant masses of the final-state particles. Of the three possible permutations, only two are independent, forming a DP for which the phase space is uniform within its boundaries.  The DP is divided into bins to gain sensitivity to the large variations in the strong-phase difference $\Delta\delta_D$ across the DP. One common binning scheme is the equal-{$\Delta\delta_D$} binning that minimises the variation in the values of $\Delta\delta_D$ in each bin; this scheme is shown in Fig.~\ref{fig:bins_KSpipi}. The DP is divided into sixteen bins, which are symmetric about the line $M^{2}_{K^0_{\text{S(L)}}\pi^+} = M^{2}_{K^0_{\text{S(L)}}\pi^-}$. The weighted averages of cosines and sines of the strong-phase difference in the $i^{\text{th}}$ bin of the DP are given by
\begin{align}
c_i &= \frac{1}{\sqrt{\int_i |A_D(\mathbf{x})|^2 \mathrm{d}\mathbf{x}
                \int_i |A_{\bar{D}}(\mathbf{x})|^2 \mathrm{d}\mathbf{x}}}
      \int_i |A_D(\mathbf{x})| |A_{\bar{D}}(\mathbf{x})|
      \cos(\Delta\delta_D) \mathrm{d}\mathbf{x}, \\
s_i &= \frac{1}{\sqrt{\int_i |A_D(\mathbf{x})|^2 \mathrm{d}\mathbf{x}
                \int_i |A_{\bar{D}}(\mathbf{x})|^2 \mathrm{d}\mathbf{x}}}
      \int_i |A_D(\mathbf{x})| |A_{\bar{D}}(\mathbf{x})|
      \sin(\Delta\delta_D) \mathrm{d}\mathbf{x},
\label{eq:cisi}
\end{align}
where $A_D(\mathbf{x})$ and $A_{\bar{D}}(\mathbf{x})$ are the decay amplitudes of $D^0$ and $\bar{D}^0$, respectively, at point 
$\text{x} = (M^2_{K_{\text{S}}^0\pi^+},\, M^2_{K_{\text{S}}^0\pi^-})$ in the same bin on the DP of the final state $K_{\text{S}}^0\pi^+\pi^-$. A similar definition of strong-phase parameters $c_i^{\prime}$ and $s_i^{\prime}$ can be written for the $K_{\text{L}}^0\pi^+\pi^-$ decay mode. 

\begin{figure}[htbp]
  \centering
  \includegraphics[width=6.5cm]{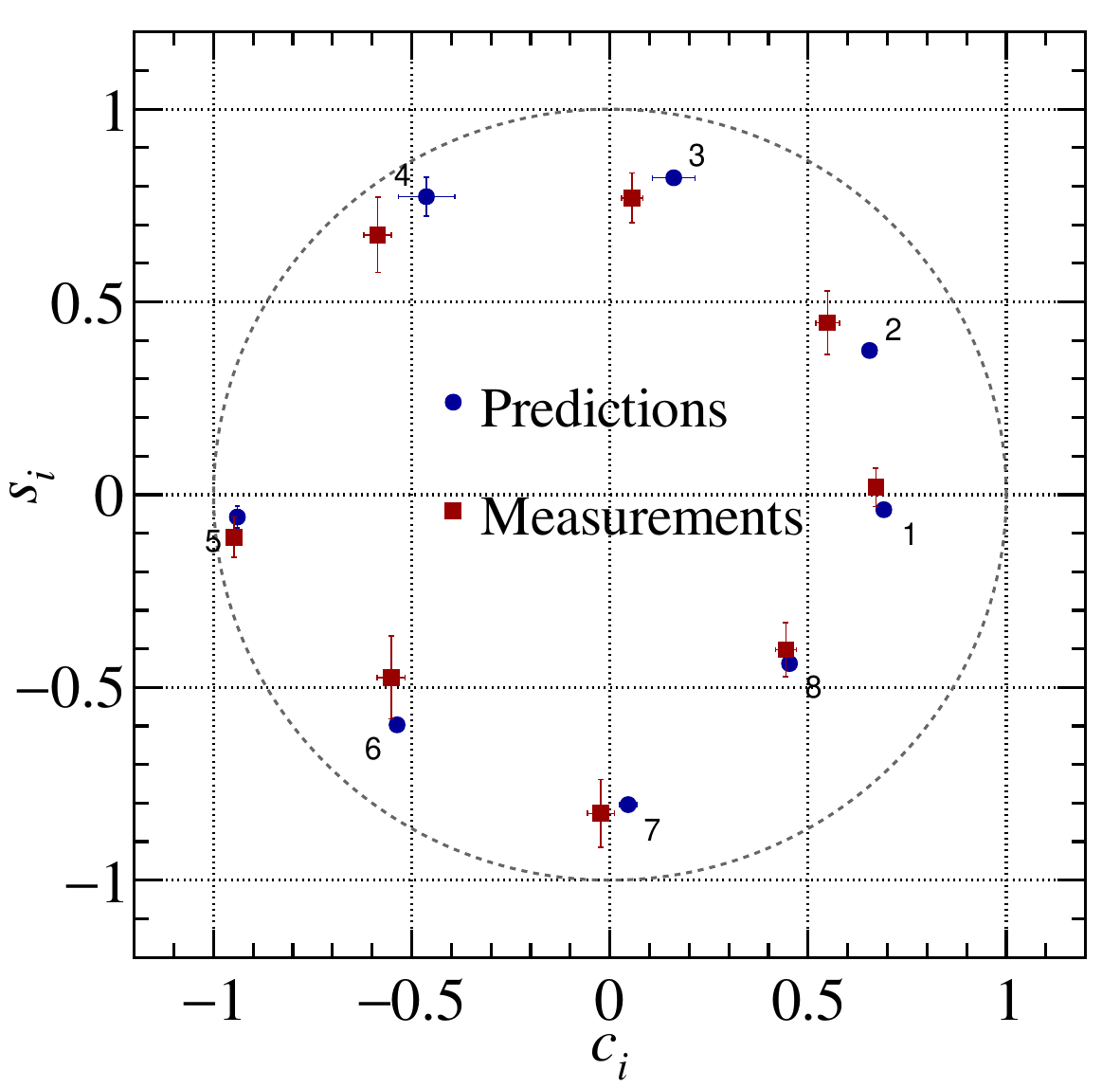}
  \includegraphics[width=6.5cm]{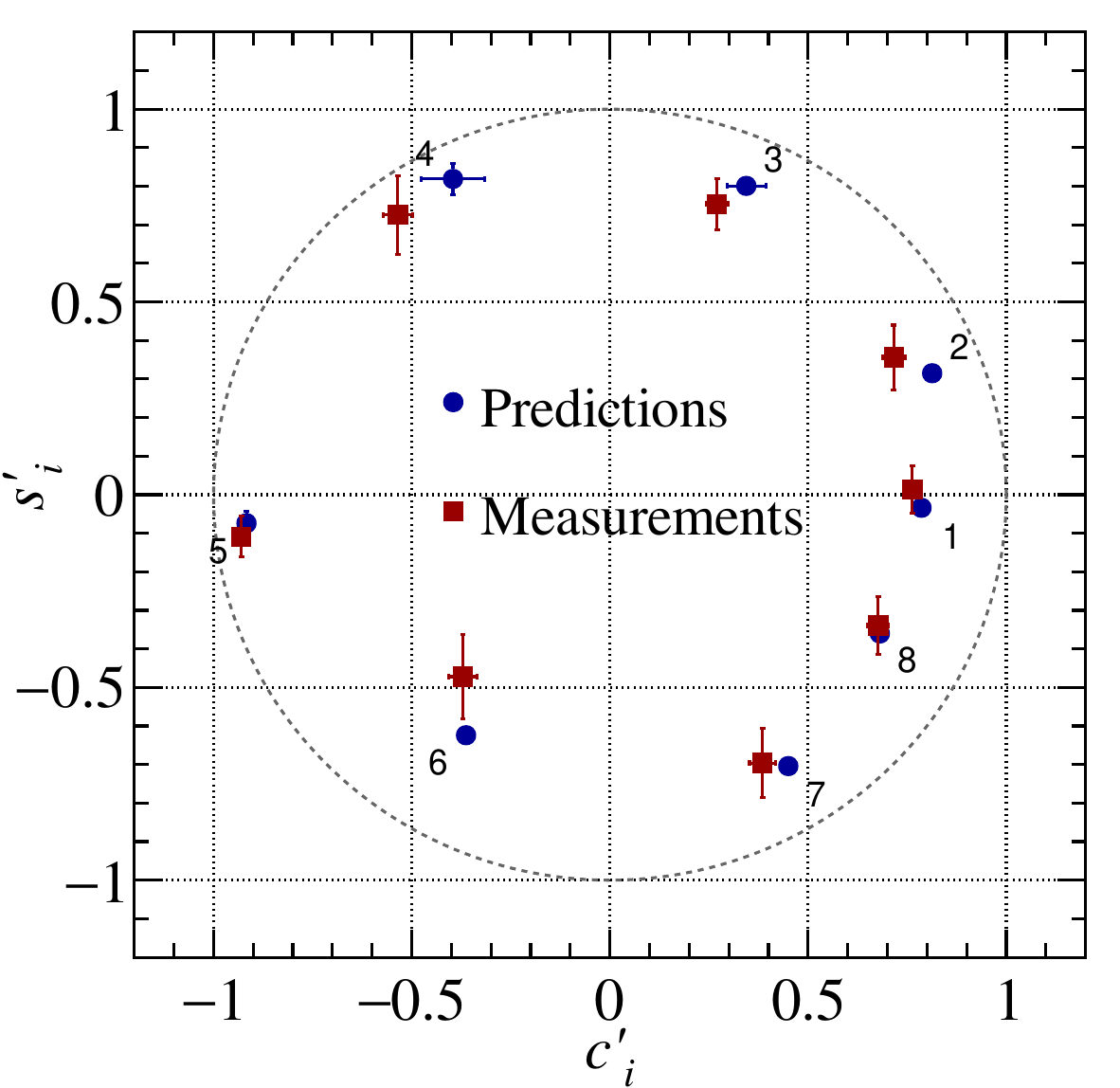}
  \caption{Model-predicted strong-phase parameters $c_i$ ($c^{\prime}_i$) and $s_i$ ($s^{\prime}_i$) are shown as blue points with error bars. Also shown for comparison are the measurements reported by the BESIII Collaboration~\cite{shenwenhan} (red points with error bars) under the equal binning scheme. The numbers 1--8 denote the different bins of the equal binning scheme.}
  \label{fig:plot_cisi}
\end{figure}

The fitted amplitude model also provides predictions for the binned strong-phase parameters 
$c_i$, $s_i$ ($K_\text{S}^0\pi^+\pi^-$) and $c_i^{\prime}$, $s_i^{\prime}$ 
($K_{\text{L}}^0\pi^+\pi^-$) under the equal binning scheme, which are listed in Table~\ref{tab:combined_cisi_values}. 
Figure~\ref{fig:plot_cisi} compares these predictions with the model-independent 
BESIII measurements~\cite{shenwenhan}. 
Overall consistency is observed, although several bins exhibit discrepancies
exceeding three standard deviations.
While the comparison involves only 16 bins, this level of discrepancy indicates that the present amplitude model does not yet fully reproduce the strong-phase structure measured directly using quantum-correlated $D\bar{D}$ data. 
Possible sources include limitations in the modelling of broad overlapping 
resonances, the treatment of the $S$-wave component, and the current
amplitude parametrisation. 
Future amplitude analyses may benefit from incorporating the
model-independent strong-phase measurements as additional constraints.

\begin{table*}[htbp]
\centering
\caption{\small Model-predicted values of the strong-phase parameters $c_i$, $s_i$, $c_i^{\prime}$ and $s_i^{\prime}$  for $D^0 \to K_\text{S}^0\pi^+\pi^-$ and $D^0 \to K_{\text{L}}^0\pi^+\pi^-$ decays. The first uncertainty is statistical, and the second is systematic.}
\resizebox{\textwidth}{!}{%
\begin{tabular}{lrrrr}
\hline
\hline
\multicolumn{1}{c}{\textbf{Bin }} & 
\multicolumn{1}{c}{\boldmath$c_i$} & 
\multicolumn{1}{c}{\boldmath$s_i$} & 
\multicolumn{1}{c}{\boldmath$c_i^{\prime}$} & 
\multicolumn{1}{c}{\boldmath$s_i^{\prime}$} \\
\hline
1 & $0.692 \pm 0.003 \pm 0.012$ & $-0.043 \pm 0.006 \pm 0.019$ & $0.787 \pm 0.002 \pm 0.010$ & $-0.036 \pm 0.005 \pm 0.011$ \\
2 & $0.657 \pm 0.004 \pm 0.016$ & $0.375 \pm 0.006 \pm 0.013$ & $0.812 \pm 0.003 \pm 0.010$ & $0.314 \pm 0.005 \pm 0.017$ \\
3 & $0.140 \pm 0.014 \pm 0.054$ & $0.828 \pm 0.003 \pm 0.008$ & $0.323 \pm 0.012 \pm 0.049$ & $0.808 \pm 0.004 \pm 0.014$ \\
4 & $-0.490 \pm 0.014 \pm 0.061$ & $0.759 \pm 0.008 \pm 0.049$ & $-0.420 \pm 0.013 \pm 0.071$ & $0.808 \pm 0.006 \pm 0.043$ \\
5 & $-0.940 \pm 0.002 \pm 0.005$ & $-0.069 \pm 0.007 \pm 0.027$ & $-0.917 \pm 0.002 \pm 0.007$ & $-0.086 \pm 0.007 \pm 0.029$ \\
6 & $-0.531 \pm 0.006 \pm 0.013$ & $-0.606 \pm 0.005 \pm 0.017$ & $-0.357 \pm 0.007 \pm 0.017$ & $-0.634 \pm 0.005 \pm 0.026$ \\
7 & $0.051 \pm 0.010 \pm 0.023$ & $-0.803 \pm 0.003 \pm 0.009$ & $0.448 \pm 0.008 \pm 0.017$ & $-0.709 \pm 0.005 \pm 0.013$ \\
8 & $0.454 \pm 0.007 \pm 0.016$ & $-0.438 \pm 0.007 \pm 0.020$ & $0.678 \pm 0.004 \pm 0.009$ & $-0.364 \pm 0.006 \pm 0.012$ \\
\hline
\hline
\end{tabular}%
}
\label{tab:combined_cisi_values}
\end{table*}

\section{Summary}\label{sec:summary}
We present an updated simultaneous amplitude analysis of $D^0 \to K_{\text{S}}^0\pi^+\pi^-$ and $D^0 \to K_{\text{L}}^0\pi^+\pi^-$ decays using the full $e^+e^-$ collision data sample corresponding to an integrated luminosity of $20.3\;\text{fb}^{-1}$ collected by the BESIII detector at $\sqrt{s}=3.773$~GeV. This work supersedes the previous BESIII analysis based on $2.93\;\text{fb}^{-1}$~\cite{BESIII:2022qvy} and benefits from significantly improved statistical precision. With the enlarged data sample, all resonance contributions included in the fit are established with significances exceeding $5\sigma$. The resulting $K_{\text{S}}^0\pi^+\pi^-$ amplitude model is consistent with the previous BABAR and Belle analyses~\cite{kpiswave}, and the numerical results agree within uncertainties. More importantly, this work substantially improves the amplitude model for $D^0 \to K_{\text{L}}^0\pi^+\pi^-$ by determining the resonance-dependent $U$-spin breaking parameters directly from data, complementing the well-established $K_{\text{S}}^0\pi^+\pi^-$ model~\cite{kpiswave}.The model reveals significant differences between the \(f_{2}(1270)\) resonant contributions to the \(K_{\text{S}}^{0}\pi^{+}\pi^{-}\) and
\(K_{\text{L}}^{0}\pi^{+}\pi^{-}\) decay amplitudes.
This demonstrates that sizeable resonance-dependent \(U\)-spin breaking is required to describe the two final states simultaneously.

The complex $U$-spin breaking parameters $\hat{\rho}$ are determined directly 
from the simultaneous fit for intermediate resonant modes that decay to the 
$CP$ eigenstate $\pi^+\pi^-$ system. 
Several measured values of $|\hat{\rho}|$ differ from unity, 
which is an assumed value. 
The resulting interference between CF and DCS 
amplitudes generates substantial asymmetries between the 
$K_\text{S}^0\pi^+\pi^-$ and $K_{\text{L}}^0\pi^+\pi^-$ FFs, 
as detailed in Sec.~\ref{sec:fitresults}.
Compared to previous analyses~\cite{cisi_1,cisi_2,cisi_3}, which assumed that $\arg(\hat{\rho})$ is uniformly distributed between $0^\circ$ and $360^\circ$, and that $|\hat{\rho}|$ follows a Gaussian distribution with mean $\tan^2\theta_C$ and width $0.5 \times \tan^2\theta_C$, the present measurement provides significantly tighter constraints on both parameters.

Model-independent strong-phase measurements from quantum-correlated $D\bar{D}$ 
data~\cite{shenwenhan} remain the primary experimental input for determinations of the 
CKM angle $\gamma$. The present analysis instead provides complementary information. 
The directly determined resonance-dependent $U$-spin breaking parameters offer quantitative insight into flavour-symmetry breaking in three-body charm decays and provide a data-driven benchmark for future theoretical and phenomenological studies.
In addition, the resulting amplitude model for $K_{\text{L}}^0\pi^+\pi^-$ may be useful in evaluating model-dependent systematic effects, such as efficiency corrections and background descriptions, in analyses involving these decay modes.
These results offer improved phenomenological input for understanding 
flavour-symmetry breaking in three-body charm decays and provide a reference 
for future theoretical and phenomenological studies. The resulting amplitude model may also prove valuable for evaluating model-dependent systematic effects, such as efficiency corrections and background descriptions, in analyses involving $D\to K_{\text{S},\text{L}}^0\pi^+\pi^-$ decays.

\clearpage

\acknowledgments
The BESIII Collaboration thanks the staff of BEPCII (https://cstr.cn/31109.02.BEPC) and the IHEP computing center for their strong support. This work is supported in part by National Key R\&D Program of China under Contracts Nos. 2023YFA1606000, 2023YFA1606704, 2025YFA1613900; National Natural Science Foundation of China (NSFC) under Contracts Nos. 11635010, 11935015, 11935016, 11935018, 12025502, 12035009, 12035013, 12061131003, 12192260, 12192261, 12192262, 12192263, 12192264, 12192265, 12221005, 12225509, 12235017, 12342502, 12361141819, 12375087, 12405112, 12447110, 12535005; the Excellent Youth Foundation of Henan Scientific Commitee under Contract No.~242300421044; the Chinese Academy of Sciences (CAS) Large-Scale Scientific Facility Program; the Strategic Priority Research Program of Chinese Academy of Sciences under Contract No. XDA0480600; CAS under Contract No. YSBR-101; 100 Talents Program of CAS; The Institute of Nuclear and Particle Physics (INPAC) and Shanghai Key Laboratory for Particle Physics and Cosmology; Agencia Nacional de Investigaci\'on y Desarrollo de Chile (ANID), Chile under Contract No. ANID CCTVal CIA250027; Istituto Nazionale di Fisica Nucleare, Italy; Knut and Alice Wallenberg Foundation under Contracts Nos. 2021.0174, 2021.0299, 2023.0315; Ministry of Development of Turkey under Contract No. DPT2006K-120470; National Research Foundation of Korea under Contract No. RS-2026-25486791; National Science and Technology fund of Mongolia; Polish National Science Centre under Contracts Nos. 2024/53/B/ST2/00975, 2025/58/E/ST2/00051; STFC (United Kingdom); Swedish Research Council under Contract No. 2019.04595; U. S. Department of Energy under Contract No. DE-FG02-05ER41374

\bibliographystyle{JHEP}
\bibliography{references}

\clearpage
\appendix

\newcommand{\BESIIIorcid}[1]{\href{https://orcid.org/#1}{\hspace*{0.1em}\raisebox{-0.45ex}{\includegraphics[width=1em]{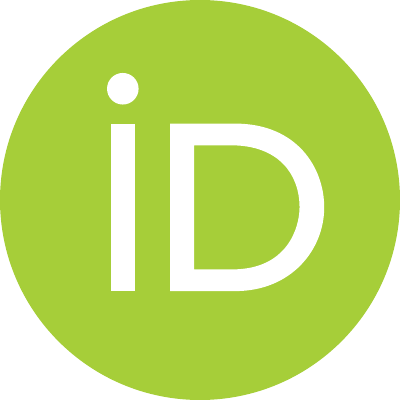}}}}
M.~Ablikim$^{1}$\BESIIIorcid{0000-0002-3935-619X},
M.~N.~Achasov$^{4,c}$\BESIIIorcid{0000-0002-9400-8622},
P.~Adlarson$^{87}$\BESIIIorcid{0000-0001-6280-3851},
X.~C.~Ai$^{93}$\BESIIIorcid{0000-0003-3856-2415},
C.~S.~Akondi$^{33A,33B}$\BESIIIorcid{0000-0001-6303-5217},
R.~Aliberti$^{41}$\BESIIIorcid{0000-0003-3500-4012},
A.~Amoroso$^{86A,86C}$\BESIIIorcid{0000-0002-3095-8610},
L.~P.~An$^{53,h}$\BESIIIorcid{0000-0002-3274-5627},
Q.~An$^{82,69,\dagger}$,
M.~S.~Anderson$^{41}$\BESIIIorcid{0009-0008-1550-2632}
% ...
\\
\vspace{0.2cm}
(BESIII Collaboration)\\
\vspace{0.2cm} {\it
$^{1}$ Institute of High Energy Physics, Beijing 100049, People's Republic of China\\
% ...
$^{4}$ Budker Institute of Nuclear Physics SB RAS (BINP), Novosibirsk 630090, Russia\\
% ...
$^{33}$ INFN Sezione di Ferrara, (A)INFN Sezione di Ferrara, I-44122, Ferrara, Italy; (B)University of Ferrara, I-44122, Ferrara, Italy\\
% ...
$^{41}$ Johannes Gutenberg University of Mainz, Johann-Joachim-Becher-Weg 45, D-55099 Mainz, Germany\\
% ...
$^{53}$ Peking University, Beijing 100871, People's Republic of China\\
% ...
$^{69}$ State Key Laboratory of Particle Detection and Electronics, Beijing 100049, Hefei 230026, People's Republic of China\\
% ...
$^{82}$ University of Science and Technology of China, Hefei 230026, People's Republic of China\\
% ...
$^{86}$ University of Turin and INFN, (A)University of Turin, I-10125, Turin, Italy; (B)University of Eastern Piedmont, I-15121, Alessandria, Italy; (C)INFN, I-10125, Turin, Italy\\
$^{87}$ Uppsala University, Box 516, SE-75120 Uppsala, Sweden\\
% ...
$^{93}$ Zhengzhou University, Zhengzhou 450001, People's Republic of China\\

\vspace{0.2cm}
$^{\dagger}$ Deceased\\
% ...
$^{c}$ Also at the Novosibirsk State University, Novosibirsk, 630090, Russia\\
% ...
$^{h}$ Also at State Key Laboratory of Nuclear Physics and Technology, Peking University, Beijing 100871, People's Republic of China\\
% ...

}

\collaborationImg{\includegraphics[width=0.15\textwidth, angle=90]{eps/AT_bes3logo.eps}}
\end{document}